\documentclass[11pt]{article}
\usepackage{amsmath,amssymb,color,cite}
\usepackage{graphicx}
\usepackage{amsfonts}
\usepackage{enumerate}
\usepackage{hyperref}
\usepackage{bbm}
\usepackage{nicefrac}
\usepackage[all]{xy}
\usepackage{graphicx}
\usepackage{booktabs}
\usepackage{amsfonts}

\def\b{{\beta}}
\def\r{{\rho}}

\def\oneone{\rlap 1\mkern4mu{\rm l}}

\newcommand{\ra}[1]{\renewcommand{\arraystretch}{#1}}
\newcommand{\hoch}[1]{$\, ^{#1}$}
\newcommand{\cN}{{\cal N}}

\makeatletter
\@addtoreset{equation}{section}
\makeatother

\newcommand{\be}{\begin{equation}}
\newcommand{\ee}{\end{equation}}
\newcommand{\bea}{\setlength\arraycolsep{2pt} \begin{eqnarray}}
\newcommand{\eea}{\end{eqnarray}}
\newcommand{\nn}{\nonumber}

\def\ft#1#2{{\textstyle{\frac{\scriptstyle #1}{\scriptstyle #2} } }}
\def\fft#1#2{{\frac{#1}{#2}}}

\def\0{{\sst{(0)}}}
\def\1{{\sst{(1)}}}
\def\2{{\sst{(2)}}}
\def\3{{\sst{(3)}}}
\def\4{{\sst{(4)}}}
\def\5{{\sst{(5)}}}
\def\6{{\sst{(6)}}}
\def\7{{\sst{(7)}}}
\def\8{{\sst{(8)}}}
\def\sst#1{{\scriptscriptstyle #1}}
\def\oneone{\rlap 1\mkern4mu{\rm l}}

\def\del{{\partial}}
\def\R{{\mathbb{R}}}

\def\im{{{\rm i}}}

\begin{document}
\begin{flushright}
\hfill{ \
MI-TH-1516\ \ \ \ }
\end{flushright}

\vspace{25pt}
\begin{center}
{\Large {\bf Holographic RG Flow in a New $SO(3)\times SO(3)$ Sector of
$\omega$-Deformed $SO(8)$ Gauged ${\cal N}=8$ Supergravity}
}

\vspace{30pt}

{\Large
Yi Pang\hoch{1}, C.N. Pope\hoch{1,2} and Junchen Rong\hoch{1}
}

\vspace{10pt}

\hoch{1} {\it George P. \& Cynthia Woods Mitchell  Institute
for Fundamental Physics and Astronomy,\\
Texas A\&M University, College Station, TX 77843, USA}

\vspace{10pt}

\hoch{2}{\it DAMTP, Centre for Mathematical Sciences,
 Cambridge University,\\  Wilberforce Road, Cambridge CB3 OWA, UK}

\vspace{20pt}

\underline{ABSTRACT}
\end{center}
\vspace{15pt}

  We consider a certain ${\cal N}=1$ supersymmetric,
$SO(3)\times SO(3)$ invariant, subsector of the
$\omega$-deformed family of $SO(8)$-gauged ${\cal N}=8$ four-dimensional
supergravities. The theory contains two scalar fields and two
pseudoscalar fields.  We look for stationary points of the
scalar potential, corresponding to AdS
vacua in the theory.  One of these, which breaks all
supersymmetries but is nonetheless stable, is new.  It exists
only when $\omega\ne 0$. We construct supersymmetric domain wall solutions in
the truncated theory, and we give a detailed analysis of their
holographic dual interpretations using the AdS/CFT correspondence.  Domain
walls where the pseudoscalars vanish were studied previously, but those
with non-vanishing pseudoscalars, which we analyse numerically, are new.
The pseudoscalars are associated with supersymmetric mass deformations in
the CFT duals.  When $\omega$ is zero, the solutions can be lifted to
M-theory, where they approach the Coulomb-branch flows of
dielectric M5-branes wrapped on $S^3$ in the deep IR.

\thispagestyle{empty}

\pagebreak
\voffset=-40pt
\setcounter{page}{1}

\tableofcontents

\addtocontents{toc}{\protect\setcounter{tocdepth}{2}}

\section{Introduction}

   For thirty years after its construction in 1982 \cite{dewitnic},
the four-dimensional $SO(8)$ gauged maximally supersymmetric
$\cN=8$ supergravity was widely considered to be a unique theory.
Interestingly, using the embedding tensor formulation \cite{deWit:2007mt},
it was recently realized that there exists a family of deformations
of the theory, characterised by a single parameter commonly
called $\omega$, associated with a mixing of the electric and
magnetic vector fields employed in
the $SO(8)$ gauging \cite{DallAgata:2012bb,dewitnicnew}. Inequivalent
$\cN=8$ theories are parameterised by values of $\omega$ in the interval
$0\le\omega\le\pi/8$.  This development has raised numerous interesting
questions,
such as its possible higher-dimensional string/M-theory origin
and the consequences of the $\omega$ deformation for the holographic
dual theory.

   The potential for the 70 scalar fields in the $\cN=8$ theory depends
non-trivially on the $\omega$ parameter, and the structure of the
stationary points, which is already rich in the original undeformed
theory, becomes even more involved in the deformed theories.  As in the
undeformed case, the investigation of the stationary points in the
complete theory is extremely complicated, and in order to render
the problem tractable, one has to consider consistent truncations in
which only subsets of the scalar fields are retained.
There have been a number of studies in which truncations of
the new $\omega$-deformed maximal supergravity have been performed, typically
with the focus being on finding scalar-field truncations in which the
scalar potential still has a non-trivial dependence on the parameter $\omega$,
leading to a richer structure of anti-de Sitter (AdS) stationary
points, with the nature of the vacuum states now being dependent on $\omega$.
One can also then look for domain-wall solutions that approach
the AdS stationary points asymptotically at infinity.

    The truncations are achieved by setting to zero all the fields
that transform non-trivially under some subgroup of the $SO(8)$ symmetry of
the original theory, thus ensuring the consistency of the truncation.
Cases that have been studied involve retaining the subset of scalar
fields invariant under an $SO(7)$, $G_2$, $SU(3)$ or
$SO(3)\times SO(3)$ subgroup \cite{DallAgata:2012bb, Borghese:2012qm, Borghese:2012zs,Borghese:2013dja, Guarino:2013gsa,Tarrio:2013qga}, or else
the seven scalars parameterising the diagonal elements of
the $SL(8,\mathbf{R})/SO(8)$ coset associated with the 35 self-dual
scalars \cite{Anabalon:2013eaa}.  These various truncations are
parallel to the consistent scalar-field truncations performed for the
the original de Wit-Nicolai theory \cite{Warner:1983vz,Warner:1983du,Ahn:2000mf,Ahn:2001kw,Ahn:2000aq,Corrado:2001nv,Bobev:2009ms,Bobev:2010ib,Cvetic:1999xx}.  Consistent
truncations retaining $U(1)$ gauge fields have also been
considered, giving rise to an $\omega$-deformed version of the
$STU$ supergravity \cite{Lu:2014fpa}, and a one-parameter extension
\cite{Cremonini:2014gia} of an Einstein-Maxwell-scalar system
\cite{Gauntlett:2009bh} previously obtained via a reduction from
eleven dimensions on a seven-dimensional Sasaki-Einstein manifold.
The latter has been used in a study of holographic condensed matter systems.

   In this paper, we consider a new consistent truncation of $SO(8)$
gauged ${\cN}=8$ supergravity,
by keeping the fields invariant under a different
$SO(3)\times SO(3)$ subgroup of $SO(8)$, which we denote by
$SO(3)_D\times SO(3)_R$. This subgroup, and its associated invariant
tensors, is defined in Appendix \ref{branching rules}.  One way to characterise
it is by starting from $SO(3)_1\times SO(3)_2\times SO(3)_3\times SO(3)_4
   \subset SO(8)$.  The factor $SO(3)_D$ is then the diagonal in
$SO(3)_1\times SO(3)_2\times SO(3)_3$, and the factor $SO(3)_R$ is
$SO(3)_4$.   The truncated theory preserves ${\cN}=1$ supersymmetry,
and it encompasses the scalar sectors invariant under $SO(7)$, $G_2$ and
$SO(4)$ as special cases.

  Within the de Wit-Nicolai theory, this sector does
not yield new critical points. However, it does provide two new critical
points in the $\omega$-deformed theories.  (These are absent in the
undeformed theory because the value of the scalar potential at these points
goes to infinity in the limit when $\omega$ goes to zero.)
By construction, the two new critical points preserve
$SO(3)_D\times SO(3)_R$ global symmetry. Moreover, one of them also
preserves ${\cN}=3$ supersymmetry in the full ${\cN}=8$ theory
\cite{Gallerati:2014xra}, while the other one, which had not been found
previously, is non-supersymmetric but nonetheless stable.

Via the AdS/CFT correspondence, stable AdS solutions in supergravity
theories correspond to local conformal field theories (CFT) living on the
boundary of AdS. Two AdS critical points may be connected by a
domain-wall solution, which is interpreted as the holographic description
of an RG flow from one CFT in the ultra-violet (UV) to another CFT in
the infra-red (IR). There are also interesting classes of holographic flows
starting from AdS in the UV and flowing to a non-AdS spacetime in the
deep IR. In such solutions, the scalar fields flow to infinite values at
the IR end of the flow, thus rendering the IR geometry singular. In fact,
most of the known domain-wall solutions belong to this class. It seems
natural to interpret these solutions as RG flows to non-conformal IR
quantum field theories.  A proper understanding of the nature of the IR
singularities of the geometry and the corresponding QFT in the IR
requires embedding the lower-dimensional solution into the UV-complete
string or M-theory. From the higher-dimensional perspective, the
singularities are physically allowable if they are associated with branes
of positive tension. Examples are holographic Coulomb-branch flows, such
as those studied in
\cite{Kraus:1998hv,Freedman:1999gk,Cvetic:1999xx}. In some other examples,
involving brane polarization \cite{Myers:1999ps}, the singularities are
placed at the locus of the dielectric branes \cite{Pope:2003jp}. A very
recent paper suggests that a singular lower-dimensional solution can lift
to a smooth higher-dimensional solution \cite{Pilch:2015vha}. Properties
of the IR QFT are also revealed by the study of the higher-dimensional
brane configuration. Depending on the nature of the sources triggering
the flow, the structure of the IR theories can take diverse forms.

In the context of the AdS$_4$/CFT$_3$ correspondence and its uplift to
M-theory, supersymmetric domain-wall solutions flowing to non-AdS spacetimes
in the IR have not been well studied and only very few examples are known.
This is a consequence of the complexity of the $\cN=8$ supergravity theory,
which contains 70 scalar fields . The main purpose of this paper is to
explore new domain-wall solutions captured by the consistently-truncated
$SO(3)_R\times SO(3)_D$-invariant sector of the $SO(8)$ gauged $\cN=8$
supergravities. We begin with the study of such supersymmetric domain-wall
solutions in the
original undeformed de Wit-Nicolai theory, since in this case embedding
within M-theory is known,  and furthermore the dual CFT is known to be the
ABJM theory. A proper interpretation in terms of a UV-complete
framework can therefore be achieved. Near the boundary AdS, the
leading fall-off coefficients of the scalars and pseudoscalars in the
truncated theory are interpreted as the vacuum expectation values (VEVs)
of dimension-1 primary operators and the supersymmetric mass terms in the
dual CFT respectively. When the pseudoscalars are turned off, the
supersymmetric domain-wall solutions were found analytically, describing
the Coulomb-branch flows on M2-branes spreading out into six possible
distributions in the transverse space. When the pseudoscalars are turned on,
the complexity of the flow equations is such that we are only able to
obtain the solutions numerically, by integrating the flow equation
from the IR to the UV. The solutions we find correspond to flows driven
by both the VEV and the mass terms. The competition between the VEV and
mass terms leads to a variety of possible IR singularities in the geometry.
The physical solutions approach the Coulomb branch flow of
dielectric M5-branes wrapping on $S^3$ in the deep IR.

   We then turn to the supersymmetric domain-wall solutions in the
$\omega$-deformed theories, within the same truncated scalar sector.
We are interested in supersymmetric holographic RG flows, and for these it
now turns out that the pseudoscalars are necessarily active.  The singular
IR behaviors of the solutions are similar to those arising in the
$\omega=0$ case. However, since the higher-dimensional origin of the
$\omega$-deformed theories is currently unknown, we must necessarily
postpone for now any attempt to give a complete interpretation of the
$\omega$-deformed supersymmetric domain-wall solutions.  In this regard,
we note that
a recent paper \cite{lescwa} contains a no-go theorem showing that the
$\omega$-deformed gauged supergravities cannot be realised via a
compactification that is locally described by ten or eleven dimensional
supergravity.

The plan of the paper is as follows.  In section 2 we discuss the
consistent truncation of the $\omega$-deformed
$\cN=8$ supergravities to the $SO(3)_D\times
SO(3)_R$ invariant sector, focusing in particular on the four scalar fields,
and their scalar potential.  We obtain the first-order equations implied by
imposing the requirement of $\cN=1$ supersymmetry on the domain-wall
solutions, and we show how the scalar potential may be written in terms
of a superpotential.  In section 3 we study the critical points of the
scalar potential.  These include a variety of
critical points that were found previously in truncations with larger
invariant symmetry groups containing $SO(3)_D\times SO(3)_R$, in addition to
the new non-supersymmetric
critical point that arises in our truncation when $\omega\ne0$.
In section 4 we discuss in detail the $\cN=1$ supersymmetric domain-wall
solutions that are asymptotic to the maximally-symmetric $\cN=8$ AdS
solution in the UV, in the case when $\omega=0$ so that we can lift the
solutions to M-theory and thus give a holographic dual interpretation
via the ABJM model.  We discuss this both for the case of vanishing
pseudoscalars, for which the domain-wall solutions had been found
analytically in earlier work, and also when the pseudoscalars are
non-vanishing, in which case we have to resort to numerical analysis.
In section 5 we extend our discussion to the case where the $\omega$
deformation parameter is non-zero.  Supersymmetric domain walls must
now necessarily have non-vanishing pseudoscalar fields, and hence all our
discussion in this section is based on the numerical analysis of the
solutions.  After presenting our conclusions in section 6, we include
two appendices in which we give details of the embedding of
$SO(3)_D\times SO(3)_R$ in $SO(8)$, and some of the conventions
for gamma matrices and uplift formulae that we employ in the paper.

\section{Truncated $\mathcal{N}=1$ SUGRA Lagrangian}

The scalar potential of the $\omega$-deformed family of
$SO(8)$ gauged $\mathcal{N}=8$ supergravities can be described conveniently
in the symmetric gauge, where the $E_7/SU(8)$ scalar coset representative
is parameterized as
\be
{\cal V} = \exp\begin{pmatrix} 0 & -\fft1{2\sqrt2} \phi_{IJKL}\cr
             -\fft1{2\sqrt2} \phi^{MNPQ} & 0\end{pmatrix}.
\ee
Here $\phi^{ijk\ell}$ are complex scalar fields, totally antisymmetric
in the rigid $SU(8)$ indices, and obeying the complex self-duality
constraint
\be
\phi_{IJKL} = \fft1{4!}\, \varepsilon_{IJKLMNPQ} \, \phi^{MNPQ}.
\ee
Note that in the symmetric gauge $SU(8)$ and $SO(8)$ indices are identified.
Introducing coordinates $x^I$ on $\R^8$ (where $I$ is an ${\bf 8}_s$ index),
the 35 complex
scalar fields can be written
as
\be
\Phi=\fft1{4!}\, \phi_{IJKL} dx^I\wedge dx^J\wedge dx^K\wedge dx^L.
\ee
The $SO(3)_D\times SO(3)_R$ invariant subset that we shall be
considering in this paper are given by
\begin{eqnarray}
\Psi_1&=& \psi_1d{x^1}\wedge d{x^2}\wedge d{x^3}\wedge d{x^8}+\bar{\psi}_1 d{x^4}\wedge d{x^5}\wedge d{x^6}\wedge d{x^7}, \nonumber\\
\Psi_2&=&\psi_2(-d{x^1}\wedge d{x^2}\wedge d{x^4}\wedge d{x^7}+d{x^1}\wedge d{x^2}\wedge d{x^5}\wedge d{x^6}+d{x^1}\wedge d{x^3}\wedge d{x^4}\wedge d{x^6}\nonumber\\&&+d{x^1}\wedge d{x^3}\wedge d{x^5}\wedge d{x^7}-d{x^2}\wedge d{x^3}\wedge d{x^4}\wedge d{x^5}+d{x^2}\wedge d{x^3}\wedge d{x^6}\wedge d{x^7})\nonumber\\
&&+\bar{\psi}_2(d{x^1}\wedge d{x^4}\wedge d{x^5}\wedge d{x^8}-d{x^1}\wedge d{x^6}\wedge d{x^7}\wedge d{x^8}+d{x^2}\wedge d{x^4}\wedge d{x^6}\wedge d{x^8}\nonumber\\&&+d{x^2}\wedge d{x^5}\wedge d{x^7}\wedge d{x^8}+d{x^3}\wedge d{x^4}\wedge d{x^7}\wedge d{x^8}-d{x^3}\wedge d{x^5}\wedge d{x^6}\wedge d{x^8}).
\nonumber\\
\label{fourform}
\end{eqnarray}
(A detailed derivation of the invariant 4-forms can be found in Appendix
\ref{branching rules}.) Here $\Psi_1$ and $\Psi_2$ parameterise
an $\frac{SL(2,R)}{SO(2)}\times \frac{SL(2,R)}{SO(2)}$ coset. Having
obtained the
form of the scalar 56-bein ${\cal V}$ for the consistent truncation we are
considering, it is a mechanical, if somewhat involved, procedure to
substitute it into the
expressions given in \cite{dewitnicnew} for the various terms in the
Lagrangian of the $\omega$-deformed $\cN = 8$
gauged supergravity.  Introducing four real scalar fields by writing
\be
\psi_1=\ft12\phi_1\, e^{\im \sigma_1},\qquad \psi_2=\ft12\phi_2\,
e^{\im \sigma_2},
\label{psidef}
\ee
the Einstein and scalar sectors of the ${\cal N}=1$ truncation are
described by the Lagrangian
\be
e^{-1}{\cal L}=R-\ft12\Big((\partial\phi_1)^2+\sinh^2\phi_1(\partial\sigma_1)^2\Big)-3\Big((\partial\phi_2)^2+\sinh^2\phi_2(\partial\sigma_2)^2\Big)-V.
\label{L1}
\ee

The potential $V$ can be written as $V=g^2 \widetilde V$, where
$g$ is the gauge coupling constant and
\bea\label{potential}
64\widetilde V &=& -256 \cosh^4\phi_2 + 2\cosh\phi_1 \cosh^2\phi_2 (
 -57 -20 \cosh 2\phi_2 + 13 \cosh 4\phi_2 + 24\sinh^4\phi_2
     \cos4 \sigma_2)\nn\\
&&+ 8\sinh\phi_1 \sinh^3 2\phi_2\, (\cos(\sigma_1-3\sigma_2) +
           3\cos(\sigma_1+\sigma_2)) \nn\\
&&+ 4\sinh^3\phi_2\, \Big\{16(\cos(2\omega-3\sigma_2) +
        3\cos(2\omega+\sigma_2))(1 -\cosh\phi_1\cosh^3\phi_2)
  \nn\\
&&
+\sinh\phi_1\sinh^3\phi_2\, (6 \sin(2\omega+\sigma_1)\sin 2\sigma_2
          -2\cos(2\omega+\sigma_1-6\sigma_2) \nn\\
&& -8 \sinh\phi_1 (3 \sinh\phi_2 + \sinh 3\phi_2)\, \cos(2\omega-\sigma_1)\nn\\
&& -\ft32 \sinh\phi_1(17\sinh\phi_2 + 5 \sinh 3\phi_2)\,
   \cos(2\omega + \sigma_1) \cos2\sigma_2 \Big\}.
\eea
Note that the potential is invariant under the transformations
\begin{eqnarray}
&&\omega\rightarrow\omega+\pi/4,\quad\sigma_1\rightarrow\sigma_1-\pi/2,\quad\sigma_2\rightarrow\sigma_2+\pi/2\label{symmetry1}\\
&&\omega\rightarrow-\omega,\quad\sigma_1\rightarrow-\sigma_1,
\quad\sigma_2\rightarrow-\sigma_2,\label{symmetry2}
\end{eqnarray}
and so inequivalent theories are characterised by the parameter
$\omega$ lying in the interval $[0,\pi/8]$.

The potential can be expressed in terms of a superpotential $W$, with
\be
\widetilde V=2\Big(4|\frac{\partial W}{\partial\phi_1}|^2+\frac2{3}|\frac{\partial W}{\partial\phi_2}|^2-3|W|^2\Big),
\ee
and
\be
W=-e^{-\im \omega}(1-|\zeta_1|^2)^{-\ft12}(1-|\zeta_2|^2)^{-3}
\Big[4\zeta_2^3e^{2\im \omega}-3\zeta_2^4-1+\zeta_1\zeta_2^6e^{2\im\omega}
+3\zeta_1\zeta_2^2e^{2\im \omega}-4\zeta_1\zeta_2^3\Big],
\ee
where we have defined
\be
\zeta_1=\tanh\ft12\phi_1 \,e^{-\im \sigma_1},
\qquad \zeta_2=\tanh\ft12\phi_2\,  e^{\im \sigma_2}.
\label{zeta}
\ee

  We are interested in ${\cal N}=1$ supersymmetric
domain-wall solutions, of the form
\be
ds^2=d\rho^2+e^{2A(\rho)}\eta_{\mu\nu}\, dx^{\mu}dx^{\nu}.
\label{dwm}
\ee
The existence of a Killing spinor requires that the first-order equations
\bea
&&\phi_1'-\im \sinh\phi_1\sigma_1'+4ge^{2\im \alpha}\frac{\partial \overline W}{\partial \phi_1}=0 , \nn\\
&&\phi_2'+\im \sinh\phi_2\sigma_2'+\frac23ge^{2\im \alpha}\frac{\partial \overline W}{\partial \phi_2}=0,\nn\\
&&A'-g|W|=0,\quad \partial_{\r}|\epsilon|-\ft12g|W|=0
\label{dweq}
\eea
should be satisfied,
where a prime denotes a derivative with respect to $\rho$ and
we use the notation $W=|W|\, e^{2\im\alpha}$.

   The solutions of these first-order equations also obey the second-order
equations of motion that follow from the Lagrangian (\ref{L1}).
This can be seen easily as follows. The action
(including the Gibbons-Hawking term) evaluated on the domain-wall
ansatz is given by
\bea
{\cal S}&=&\int d\r \, e^{3A}\Big[6(A'-g|W|)^2-\ft12|\phi_1'-
  \im \sinh\phi_1\sigma_1'+
4g\frac{\partial \bar W}{\partial \phi_1}e^{2\im \alpha}|^2\nn\\
&&\qquad-3|\phi_2'+\im \sinh\phi_2\sigma_2'+\frac23g
\frac{\partial \bar W}{\partial \phi_2}e^{2\im \alpha}|^2\Big]+
\Big[4g\,e^{3A}\,|W|\Big]^{\infty}_{-\infty},
\eea
and this is clearly extremised by the solutions of (\ref{dweq}).
(Here we adopt the same notation as \cite{Ahn:2000mf}.)
Using $\gamma_2\,\epsilon_8=(\epsilon^{8})^*$, and
\be
\partial_{\sigma_1}|W|=|W|\sinh\phi_1\partial_{\phi_1}\arg W,\qquad \partial_{\sigma_2}|W|=-|W|\sinh\phi_2\partial_{\phi_2}\arg W ,
\label{wid}
\ee
the solutions to the Killing spinor equation are given by
\be
\epsilon_8=e^{\ft12A(\rho)+\im \alpha}\left(
                                   \begin{array}{c}
                                     \varepsilon_1 \\
                                     \varepsilon_2 \\
                                      0\\
                                      0\\
                                   \end{array}
                                 \right),
\ee
where $\varepsilon_1$ and $\varepsilon_1$ are two real constants.
Utilizing (\ref{wid}), eqs (\ref{dweq}) can be rewritten as
\be
\phi^{I\prime}= -2g\, {\cal K}^{IJ}\,
\frac{\partial|W|}{\partial\phi^J},\qquad A'=g\, |W|,
\label{gradiflow}
\ee
where $\phi^I$ denotes all four real scalars, and ${\cal K}^{IJ}$ is the
inverse metric for the kinetic terms in the scalar coset.
Eq.~(\ref{gradiflow}) implies that the BPS equations in the bulk describe
a gradient flow in the scalar coset manifold, with $-g|W|$ being
the ``potential'' whose gradient drives the flow. When the solutions
to (\ref{gradiflow}) are asymptotically-AdS domain walls and therefore
correspond to an RG flow in the dual CFT, Eq.~(\ref{gradiflow})
also implies a holographic strong $a$-theorem\footnote{Recently,
gradient flows and a strong $a$-theorem were studied in three
dimensions \cite{Jack:2015tka}, demonstrating the existence of a
candidate $a$-function for renormalisable Chern-Simons matter theories
at two-loop order. The monotonic behavior of the $a$-function along
renormalisation group flows is related to the $\beta$-function
via a gradient flow equation involving a positive-definite metric similar
to that in our holographic discussion. The function $A$ defined in
\cite{Jack:2015tka} is equivalent to our  $|W|$ here.}
\be
A''=-2g^2 \, \frac{\partial|W|}{\partial\phi^I}\, {\cal K}^{IJ}\,
\frac{\partial|W|}{\partial\phi^J}\leq0.
\ee

\section{Critical Points of the Scalar Potential}

\begin{figure}[h]
\includegraphics[width=6cm]{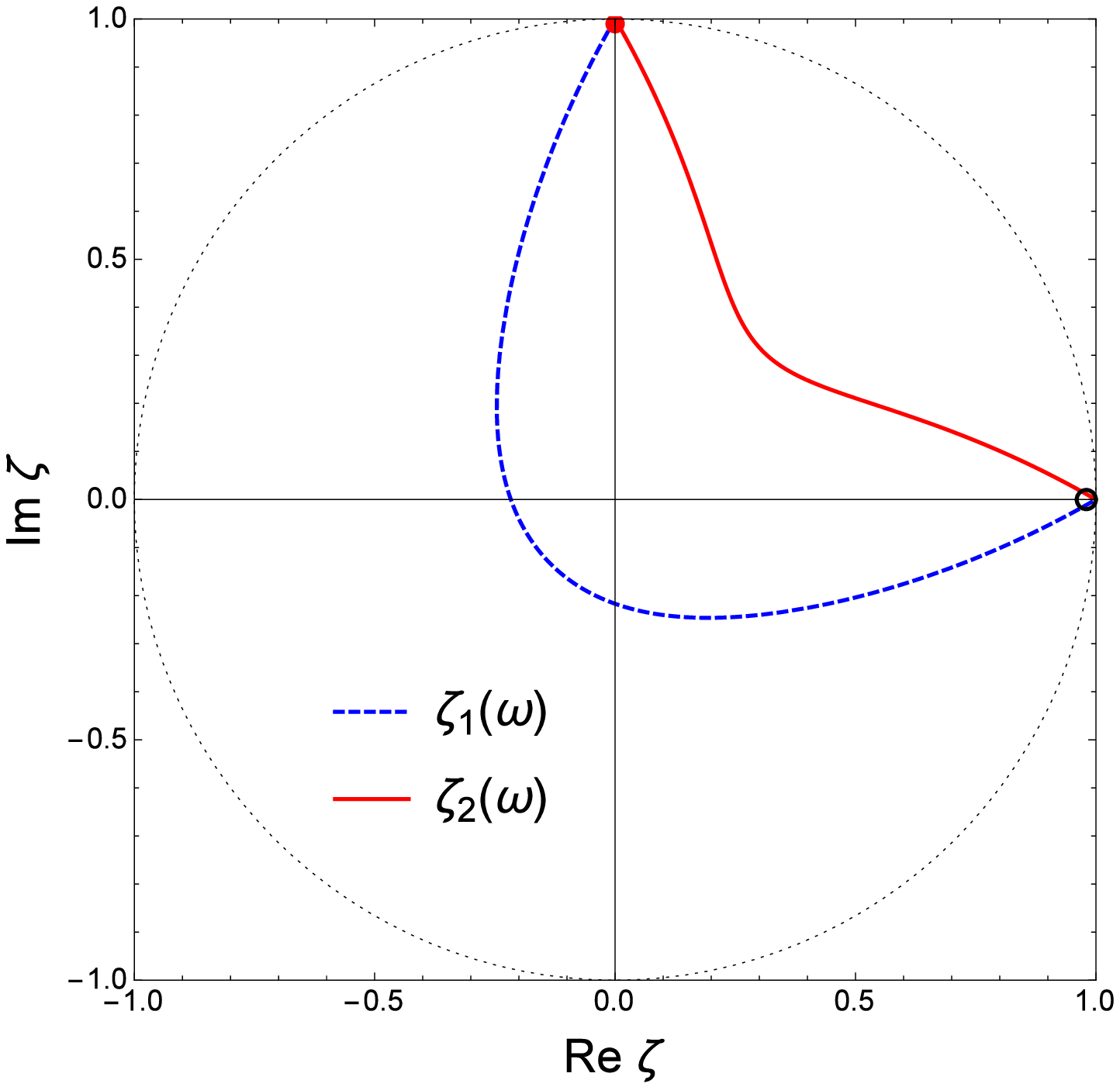}
\includegraphics[width=8.8cm]{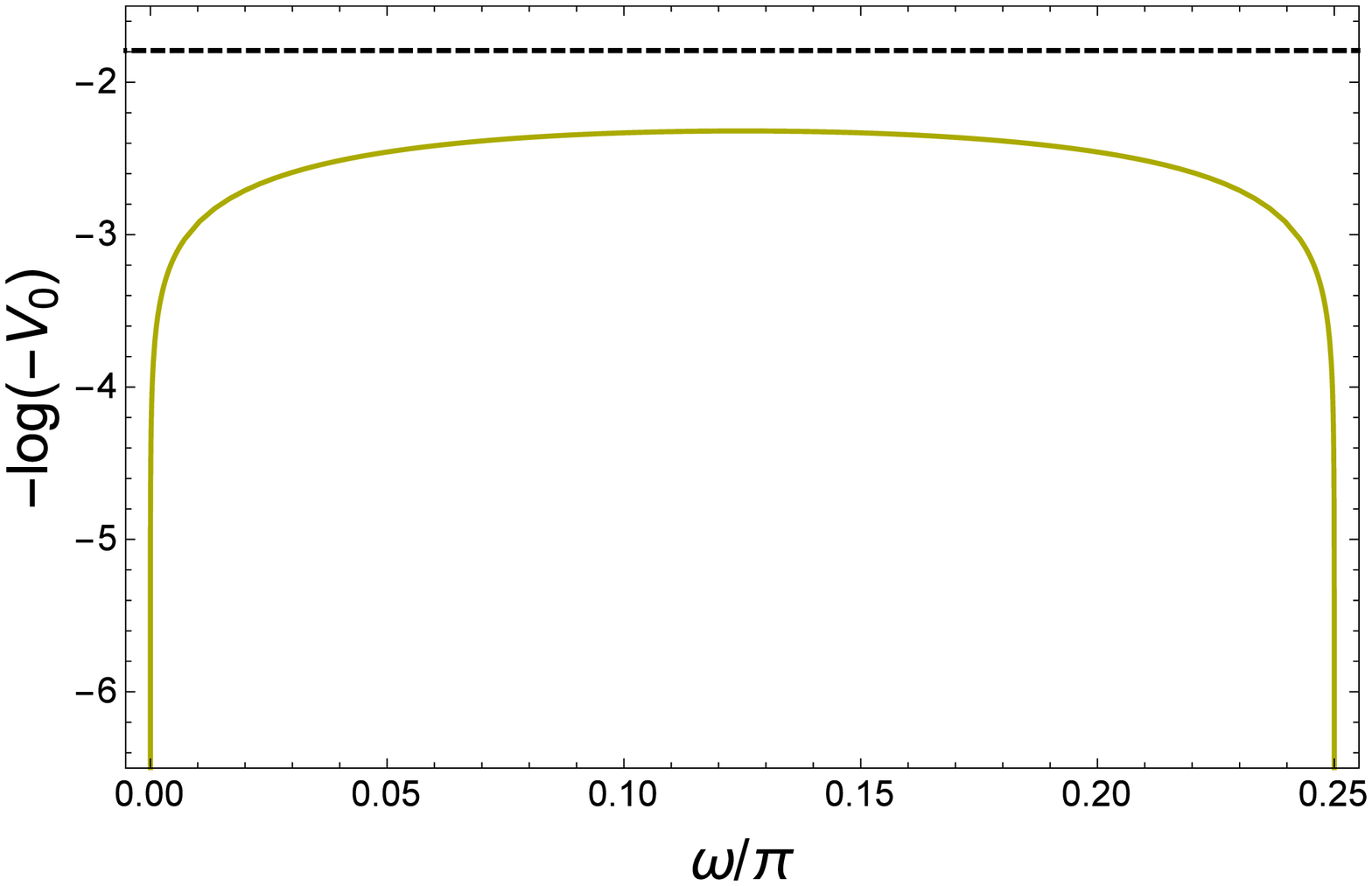}
\caption{\it $\omega$ dependence of $\cN=3$ $SO(3)_D\times SO(3)_R$
critical point. The black circle (on the right of the plot) is $\omega=0$;
the red dot (at the top) is $\omega=\pi/4$.
The dashed line in the right-hand plot corresponds to $V_0=-6$. }\label{point1}
\end{figure}

The symmetry group $SO(3)_D\times SO(3)_R$ can be embedded in $SO(8)$ through
the chain
\be
SO(3)_D\times SO(3)_R\subset G_2\subset SO(7)\subset SO(8).
\ee
On the other hand the $SU(3)$ invariant sector of $\cN=8$ supergravity has been thoroughly studied both in the original de Wit and Nicolai theory\cite{Warner:1983du,Warner:1983vz,Bobev:2010ib} and in the $\omega$-deformed case\cite{Borghese:2012zs}, with the group embedding
\be
SU(3)\subset G_2\subset SO(7)\subset SO(8).
\ee
Using the Newton-Raphson method, with the potential \eqref{potential},
we scanned for its critical points. We found all the previously-known
critical points with $G_2$ or $SO(7)$ symmetry, and also
two critical points with $SO(3)_D\times SO(3)_R$ symmetry.
One of them, preserving $\cN=3$ supersymmetry, was first discovered in
\cite{Gallerati:2014xra}. For this critical point, the dependence of the
two complex scalars, and the associated cosmological constant, on $\omega$
are displayed in Fig. \ref{point1}. This was the first example of an
$\cN=3$ supersymmetric vacuum in $SO(8)$ gauged $\cN=8$ supergravity.
The mass spectrum of the fluctuations around this vacuum is given by
\cite{Gallerati:2014xra}:
\begin{eqnarray}
m^2 L_0^2 & : & \quad 1\times(3(1+\sqrt{3})); \quad
 6\times(1+\sqrt{3});\quad 1\times(3(1-\sqrt{3}));\quad
 6\times(1-\sqrt{3});\nonumber\\
&&\quad 4\times(-\frac{9}{4});\quad
 18\times(-2);\quad 12\times(-\frac{5}{4});22\times 0 .
\end{eqnarray}
(The integer to the left of the multiplication sign indicates the
degeneracy of the mass eigenvalue, while the number to the right indicates
the corresponding mass-squared.)
Owing to the supersymmetry, the Breitenlohner-Freedman bound $m^2\, L_0^2 \ge
-\ft94$ \cite{Breitenlohner:1982jf} is necessarily respected.

\begin{figure}[h]
\includegraphics[width=6cm]{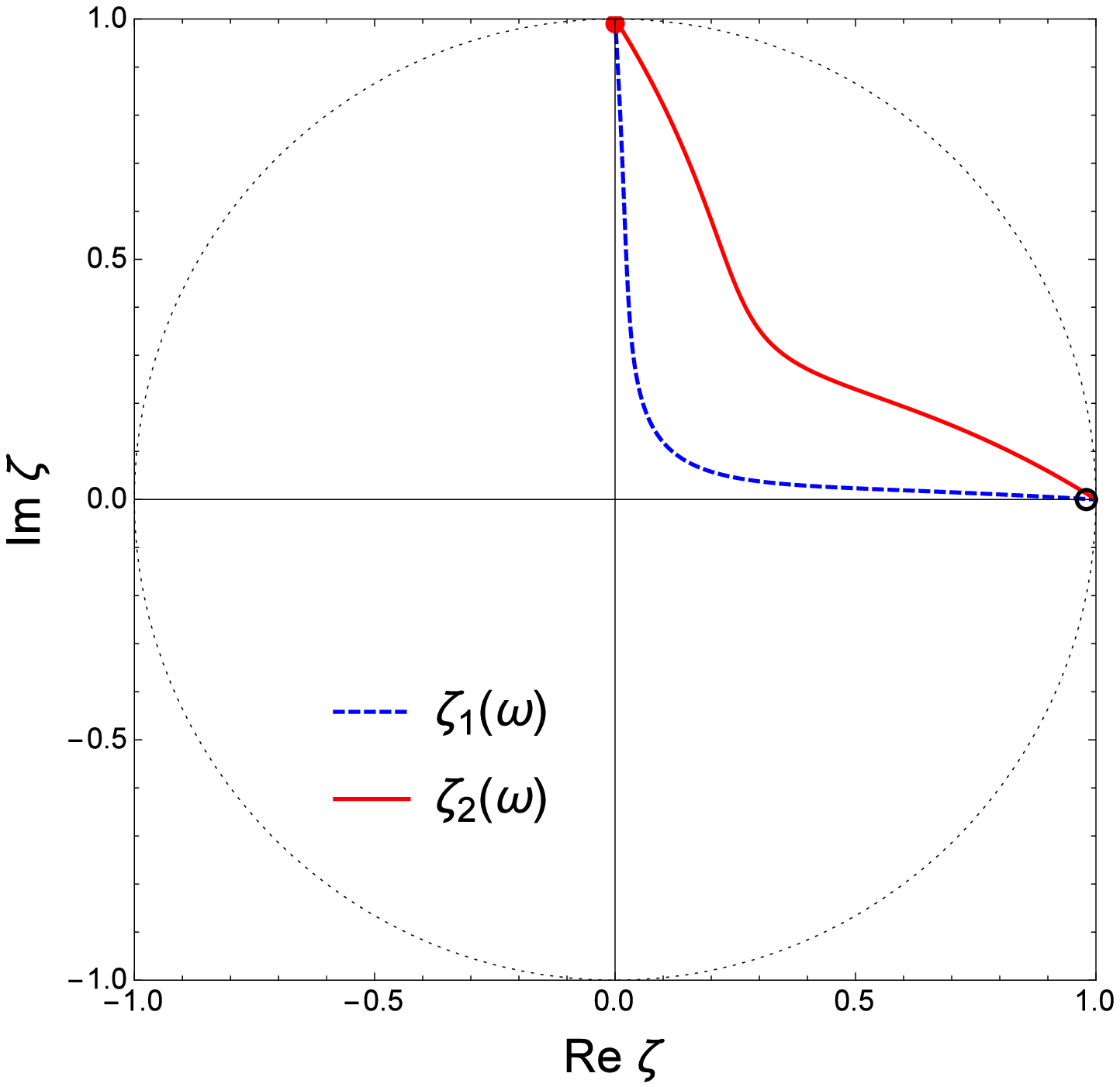}
\includegraphics[width=8.8cm]{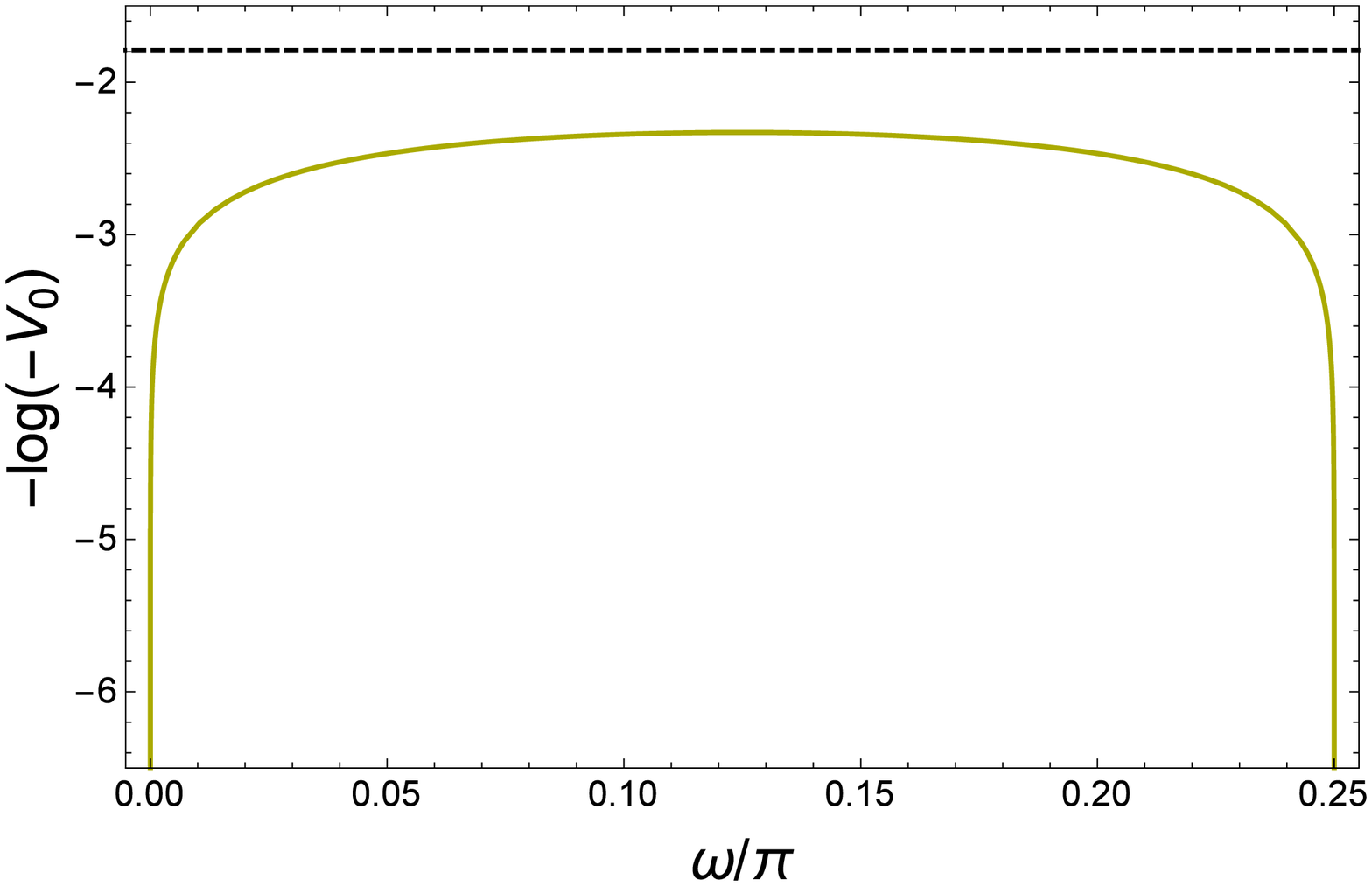}
\caption{\it{ $\omega$ dependence of the $\cN=0$ $SO(3)_D\times SO(3)_R$
critical point.  The black circle (on the right of the plot) is $\omega=0$; the
red dot (at the top) is $\omega=\pi/4$. The dashed
line in the right-hand plot corresponds to $V_0=-6$.} }\label{point2}
\end{figure}

The other $SO(3)_D\times SO(3)_R$ critical point that we found
is non-supersymmetric.  It is, however, stable against fluctuations.  The mass
spectrum of the perturbations around this vacuum depends on the value
of $\omega$.
For example, for $\omega=\pi/8$ the spectrum is given by
\begin{eqnarray}
m^2 L_0^2 & : & \quad 1\times(6.72079);\quad 1\times(5.29013);
\quad 4\times(-1.96647);\quad 9\times(-1.73861);\nonumber\\
&&\quad 9\times(-1.60284);\quad 1\times(-1.59124);8\times(-1.18046);
\quad 5\times(-0.98076);\nonumber\\
&&\quad 4\times(-0.73134);\quad 5\times(0.61746);\quad
1\times(0.58185);\quad 22\times 0.
\end{eqnarray}
Note that even though it is not supersymmetric, the Breitenlohner-Freedman
bound is not violated. For this critical point, we show in the left-hand
plot of Fig. \ref{point2} the evolution of the values of the
scalars as $\omega$ varies from 0 to $\pi/4$. The right-hand plot shows
the dependence of the cosmological constant on $\omega$. It should be
noted that for each of the $SO(3)_D\times SO(3)_R$ vacua, the
critical point disappears when
$\omega$ goes to zero, since the value of the cosmological constant
then diverges in each case.
For completeness, in Table \ref{criticalpoints}, we list all the critical points contained in the $SO(3)_D\times SO(3)_R$ invariant sector.

Three remarks are in order:
\\a) The two transformations \eqref{symmetry1} and \eqref{symmetry2}
combine into a symmetry in the case that $\omega=\pi/8$:
\begin{equation}
\sigma_1\rightarrow-\sigma_1-\pi/2,\quad\sigma_2\rightarrow-\sigma_2+\pi/2.
\end{equation}
This can be seen in the table.
\\b) Points related by
$$\phi_i\rightarrow-\phi_i,\quad\sigma_i\rightarrow\sigma_i+\pi$$ have the same location in the complex plane, and hence are
equivalent.
\\c) It is interesting to see that there are two critical points with the same cosmological constant, but with different residual symmetry; one with
$G_2$ and the other with $SO(3)_D\times SO(3)_R$ invariance.

\begin{table*}\centering
\ra{1.1}
{\small
\begin{tabular}{@{}rrrcrrcrrcrr@{}}
\toprule
 \midrule
symmetry&{$\phi_1$} && {$\sigma_1$} &&{$\phi_2$} &&
{$\sigma_2$} &&{$V (g=1)$}&{stability}
 \\ \midrule
$\cN=8$\\$SO(8)$
 &0 && $-$ && $0$ && $-$  && $-6$ & $\surd$\\ \midrule
$\cN=0$\\ $SO(7)_-$
  &0.4195&& $\ft{\pi}2$  && $0.4195$ && $-\ft{\pi}2$  && $-6.7482$ & $\times$\\ \midrule
$\cN=0$\\$SO(7)_-$
  &0.6406&& $-\ft{\pi}2$  && $0.6406$ && $\ft{\pi}2$  && $-7.7705$ & $\times$\\ \midrule
$\cN=0$\\ $SO(7)_+$
  &0.4195&& $\pi$  && $0.4195$ && $\pi$  && $-6.7482$ & $\times$\\ \midrule
${}^*\cN=0$\\ $SO(7)_+$
  &0.6406&& $0$  && $0.6406$ && $0$  && $-7.7705$ & $\times$\\ \midrule
$\cN=1$\\ $G_2$
  &0.4840&& $\ft{3\pi}4$  && $0.4195$ && $\ft{3\pi}4$  && $-7.0397$ & $\surd$\\\midrule
$\cN=1$\\ $G_2$
  &0.6579&& $-1.9693$  && $0.6579$ && $1.9693$  && $-7.9430$ & $\surd$\\\midrule
${}^*\cN=1$\\ $G_2$
  &0.6579&& $0.3985$  && $0.6579$ && $-0.3985$  && $-7.9430$ & $\surd$\\\midrule
${}^*\cN=0$\\ $G_2$
  &$\log \vartheta$&& $-\ft{\pi}4$  && $\log \vartheta$ && $\ft{\pi}4$  && $-4\vartheta$ & $\surd$\\\midrule
${}^*\cN=3$\\ $SO(3)_D\times SO(3)_R$
  &$\log \vartheta/\sqrt{3}$&& $\ft{3\pi}4$  && $\log \vartheta$ && $\ft{\pi}4$  && $-4\vartheta$ & $\surd$\\\midrule
${}^*\cN=0$\\ $SO(3)_D\times SO(3)_R$
  &$0.3114$&& $-\ft{\pi}4$  && $0.9914$ && $\ft{\pi}4$  && $-10.271$ & $\surd$\\\midrule
 \bottomrule
\end{tabular}
}\normalsize
\caption{\it Critical points in the $\omega=\pi/8$ theory.
We use $\vartheta$ to denote the number $\sqrt{3+2\sqrt{3}}$.
The mass spectra of fluctuations around the critical points are
independent of $\omega$, except for the last one. Points marked
with ``*'' disappear when $\omega\rightarrow0$, while the two points
with $SO(7)_-$ symmetry become degenerate in energy.}
\label{criticalpoints}
\end{table*}

\section{Holographic ${\cal N}=1$ RG Flows on M2-branes }

In this section, we study the domain-wall solutions to Eq.\eqref{dweq}. In
particular, we are interested in solutions approaching the
trivial ${\cal N}=8$ AdS vacuum in the UV. We shall first restrict our
discussion to the $\omega=0$ case, in which the supergravity is just
the original de Wit-Nicolai theory.  The boundary CFT corresponding to the
trivial ${\cal N}=8$ vacuum is the ABJM theory \cite{Aharony:2008ug}
with Chern-Simons level
$k=1$ or $k=2$, for which the supersymmetry is enhanced from ${\cal N}=6$
to ${\cal N}=8$ \cite{Benna:2009xd,Gustavsson:2009pm,Kwon:2009ar}. Therefore, our domain-wall solutions describe
the RG flows on M2-branes driven by ${\cal N}=1$ deformations.

  Owing to the fact that $SO(8)$ is manifest in the gravity theory but
is not manifest in the ABJM theory under the large $N$ limit, it is not
straightforward to map the bulk scalars to the boundary primary operators.
However, we recall that the $SU(2)\times SU(2)$ ABJM theory is equivalent
to the BLG \cite{Bagger:2006sk, Gustavsson:2007vu} theory,
which {\it is} manifestly $SO(8)$ invariant. Thus, to infer the form of
the primary operators dual to the $SO(3)_{ D}\times SO(3)_{ R}$
invariant bulk scalar fields, we make the reasonable assumption that
the structures of the primary operators characterised by representations
of
the R-symmetry group do not depend explicitly on the rank of the gauge group.
Based on this assumption, we identify the dual primary operators in
the large-$N$ ABJM theory by first mapping the bulk scalars to
primary operators in
the BLG theory, and then recasting these operators in terms of the fields
in the $SU(2)\times SU(2)$ ABJM theory.

We divide the 70 bulk scalars into real (self-dual) and imaginary
(anti-self dual) parts, by writing
\be
\phi_{ijkl}=S_{ijkl}+\im P_{ijkl}.
\ee
The 70 scalars can be mapped into two 35-dimensional symmetric traceless tensors as follows:
\be
{\cal O}^{(1)}_{IJ}=(\Gamma^{ijkl})_{IJ}\, S_{ijkl},\qquad
{\cal O}^{(2)}_{\alpha\beta}=(\Gamma^{ijkl})_{\alpha\beta}\, P_{ijkl},
\label{map}
\ee
where $(i, I, \alpha)$ label the $({\bf 8}_c, {\bf 8}_v, {\bf 8}_s)$
representations of $SO(8)$ respectively. The $SO(8)$ gamma matrices
are expressible in terms of triality rotation matrices. In our conventions,
the $SO(8)$ gamma matrices take the form
\be
\Gamma^i=\left(\begin{array}{cc}
0& \hat{\Gamma}^{i}_{I\alpha}\\
(\hat{\Gamma}^i)^{\rm T}_{\alpha I} & 0 \\
\end{array}\right).
\ee
The details of the $SO(8)$ gamma matrices can be found in Appendix B.

 From now on, for simplicity of discussion, we shall without loss of generality
choose the gauge coupling
constant to be $g=1$. Consequently, the trivial ${\cal N}=8$ vacuum is
the standard AdS spacetime with unit radius, and the perturbations of the
70 scalars around this vacuum have mass squared $m^2=-2$. In terms of
the coordinate $z=e^{-\rho}$, the boundary of AdS is defined at
$z\rightarrow 0$. Near the boundary of the ${\cal N}=8$ supersymmetric
AdS vacuum, the scalars $\phi_{ijkl}$ behave as
$\phi_{ijkl}\sim z\, \phi^{(1)}_{ijkl}+z^2\, \phi^{(2)}_{ijkl}$ in which the two modes are both renormalizable \cite{Klebanov:1999tb}. Boundary
conditions preserving ${\cal N}=8$ supersymmetry require
\cite{Hawking:1983mx,Borghese:2014gfa}
\be
S_{ijkl}\sim z\, S_{ijkl}^{(1)}, \qquad P_{ijkl}\sim z^2\, P^{(2)}_{ijkl}.
\ee
This set of boundary conditions amounts, in the dual holographic picture,
to $S_{ijkl}^{(1)}$ and $P^{(2)}_{ijkl}$ being the VEVs of
35 dimension-1 scalar operators and 35 dimension-2 pseudoscalar
operators in the dual SCFT respectively. In terms of the fields in the
BLG theory, these operators take the form
\be
{\bf 35}_v: {\rm Tr}(\phi_I\phi_J)-\ft18\delta_{IJ}{\rm Tr}(\phi_K\phi_K),
\qquad {\bf 35}_s: {\rm Tr}(\psi_{\alpha}\psi_{\beta})-\ft18\delta_{\alpha\beta}{\rm Tr}(\psi_\lambda\psi_\lambda).
\label{ops}
\ee
The expansion coefficients $S_{ijkl}^{(2)}$ and $P^{(1)}_{ijkl}$ then
correspond to the sources for these operators, according to the
standard AdS/CFT dictionary. However, we shall see below that besides
the source terms, the coefficients $S_{ijkl}^{(2)}$ can depend on terms
quadratic in the VEVs of the ${\bf 35}_v$ operators. This phenomenon
has been seen  previously, in the  study of a continuous distribution
of branes (see \cite{Gubser:2000nd} for a review). As far as we are aware,
there is no clear way to related these terms to the VEVs of
higher-dimension operators. Also, we emphasize that when the dual
SCFT is deformed by some operators, the operators in \eqref{ops} will
receive corrections.

We consider the domain-wall solutions asymptotic to the trivial ${\cal N}=8$ vacuum in the UV at $z=0$.
Denoting $\zeta_1(z)=S(z)+\im P(z)$ and
$\zeta_2(z)=\widetilde{S}(z)+\im\widetilde{P}(z)$, we find that near the
UV boundary, the solutions to Eq.~\eqref{dweq} take the form
\bea
S(z)&=& \alpha_1\, z+\alpha_2\, z^2+\dots ,\nonumber\\
P(z)&=& \beta_1\, z+\beta_2\, z^2+\dots ,\nonumber\\
\widetilde{S}(z)&=& \widetilde{\alpha}_1\, z
   +\widetilde{\alpha}_2\, z^2+\dots ,\nonumber\\
\widetilde{P}(z)&=& \widetilde{\beta}_1\, z+\widetilde{\beta}_2\, z^2+\dots,
\label{boundaryseries}
\eea
where
\be
\alpha_2=-3\widetilde{\alpha}_1^2+3\widetilde{\beta}_1^2,\qquad \widetilde{\alpha}_2=-\alpha_1 \widetilde{\alpha}_1-2 \widetilde{\alpha}_1^2+\beta_1\widetilde{\beta}_1+2\widetilde{\beta}_1^2,
\label{bosonmass1}
\ee
and
\be
\beta_2=6\widetilde{\alpha}_1\widetilde{\beta}_1,\qquad \widetilde{\beta}_2=\widetilde{\alpha}_1\beta_1+\alpha_1\widetilde{\beta}_1+4\widetilde{\alpha}_1\widetilde{\beta}_1.
\label{fermionvev}
\ee
(Recall that $\zeta_1$ and $\zeta_2$ parameterise the 4-form
\eqref{fourform}.) Using \eqref{map}, the 4-form is mapped to
\bea
{\cal O}^{(1)}_{IJ}&=&S\, {\rm diag}(-1,-1,-1,-1,1,1,1,1)
   +\widetilde{S}\, {\rm diag}(0,0,0,0,-2,-2,-2,6),\nn\\
{\cal O}^{(2)}_{\alpha\beta}&=&P\, {\rm diag}(-1,-1,-1,-1,1,1,1,1)
   +\widetilde{P}\, {\rm diag}(0,0,0,0,-2,-2,-2,6).
\label{dualop}
\eea
In general, ${\cal O}^{(1)}$ and ${\cal O}^{(2)}$ will differ from each
other by a similarity transformation generated by $\Gamma_{I\alpha}^8$,
which is the identity matrix in our conventions.
The expressions \eqref{dualop} suggest that the ${\bf 35}_v$ operators
receive VEVs proportional to ${\cal O}^{(1)}$, and that also
a fermion mass term
of the form ${\cal O}^{(2)}$ is turned on in the dual SCFT. The
supersymmetric completion of the fermion mass term must include
a bosonic mass term whose coefficient should be quadratic in the fermionic mass
parameter. This implies that the quadratic $\beta$ terms in
\eqref{bosonmass1} are related to the boson mass parameter, since
$\beta_1$ and $\beta_2$ correspond to the fermion mass parameters.
We can check the relation between the boson and fermion mass parameters
explicitly, by utilizing the ${\cal N}=1$ formulation of the
BLG theory \cite{Mauri:2008ai}. As mentioned previously, the gauge group
of the BLG theory is $SU(2)\times SU(2)$, and therefore one cannot take a
large-$N$ limit of the theory with $N$ being the rank of the gauge group.
Nonetheless, we shall see that a naive application of the
BLG theory gives rise to a result matching with the gravity dual.

The ${\cal N}=1$ formulation of the BLG theory is expressed in terms of
the superfield
\begin{equation}
\Phi_I=\phi_I+\theta\,\hat{\Gamma}^{8}_{I \alpha}\,
\psi^{\alpha}- \theta^2\,  F_I,
\label{superfield}
\end{equation}
where the gauge group indices and the $SO(1,2)$ spinor indices are suppressed. In order to have the fermion
mass term of the form ${\cal O}^{(2)}$ in \eqref{dualop}, the BLG action
must be deformed by
\be
\Delta {\cal L}_{\rm BLG} = {\cal O}^{(2)}_{IJ} \, {\rm Tr}(\Phi_I\Phi_J),
\ee
since in our conventions $\hat{\Gamma}^{8}_{I \alpha}=\delta_{I\alpha}$.
In the component language, this deformation contains terms of the form
${\cal O}^{(2)}_{IJ}\,{\rm Tr}(\phi_I F_J)$. The supersymmetric kinetic
term contains a contribution $\ft12{\rm Tr}(F_IF_I)$ quadratic in the auxiliary
field $F_I$. Integrating out $F_I$ generates a positive-definite mass
term $\ft12 ({\cal O}^{(2)})^2_{IJ}\, \phi_I\phi_J$ for $\phi_I$.
Decomposing the mass matrix into its traceless part and its trace, we find that the traceless part is given by
\be
\ft12 ({\cal O}^{(2)})^2_{\{IJ\}}=3\widetilde{\beta}^2_1\,
   {\rm diag}(-1,-1,-1,-1,1,1,1,1)
+(\beta_1\widetilde{\beta}_1+2\widetilde{\beta}_1^2)\,
    {\rm diag}(0,0,0,0,-2,-2,-2,6),
\label{bosonmass2}
\ee
whilst the trace part takes the form
\be
\ft12 ({\cal O}^{(2)})^2_{IJ}\,\delta^{IJ}
  =\ft12(\beta_1^2+6\widetilde{\beta}_1^2).
\ee
From Eq.\eqref{bosonmass2}, we see the familiar $8\times 8$
matrices associated with the $SO(3)_{\rm D}\times SO(3)_{\rm R}$
invariant scalars. Miraculously, the coefficients in front of these
matrices are exactly the same as the quadratic $\beta$ terms in the
sub-leading expansion coefficients of the self-dual scalars $S$ and
$\widetilde{S}$. In the standard AdS/CFT dictionary, the trace part of
the boson mass matrix is not captured by supergravity fields. The
associated operator ${\rm Tr}(\phi_I\phi_I)$ belongs to the Konishi
multiplet which is dual to the shortest stringy mode. However, as we have
seen, the non-chiral operator ${\rm Tr}(\phi_I\phi_I)$ appears in the
${\cal N}=1$ mass terms. In fact, there is no contradiction. As pointed
out in \cite{Aharony:1999ti}, when the CFT is deformed the form of
the chiral operators changes, and they mix with other operators. In
the case of the supersymmetric mass deformation, the chiral
operator \eqref{ops} mixes with the non-chiral operator
${\rm Tr}(\phi_I\phi_I)$,  giving the scalars a positive-definite mass.

In the following, we briefly discuss how to reformulate the ${\bf 35}_v$
and ${\bf 35}_s$ operators in the framework of the ABJM theory.
The global symmetry of the ABJM theory is $SU(4)\times U(1)_{\rm b}$, rather
than $SO(8)$. Under this $SU(4)\times U(1)_{\rm b}$ subgroup,
the ${\bf 8}_c$ of $SO(8)$ branches into
\be
{\bf 8}_c\rightarrow {\bf 6}_{0}+{\bf 1}_{2}+{\bf 1}_{-2},
\ee
which implies
\bea
{\bf 8}_v\rightarrow {\bf 4}_{1}+\bar{{\bf 4}}_{-1},&&\qquad
{\bf 8}_s\rightarrow {\bf 4}_{1}+\bar{{\bf 4}}_{-1},\nn\\
{\bf 35}_v \rightarrow {\bf 10}_{2}+\bar{{\bf 10}}_{-2}+{\bf 15}_0,
&&\quad {\bf 35}_s \rightarrow {\bf 10}_{-2}+\bar{{\bf 10}}_{2}+{\bf 15}_0.
\label{ABJM branching}
\eea
Group theoretically, \eqref{ABJM branching} means that the ${\bf 35}_v$
operators in \eqref{ops} should be replaced by three sets of operators in
the ABJM theory, namely
\bea
{\bf 15}_0:&&\quad  {\rm Tr}\left(Z^{\dagger}_AZ^B-
   \frac{1}{4}\delta^B_A Z^{\dagger}_C Z^C\right),\nn\\
{\bf 10}_2:&&\quad  {\rm Tr}\left(Z^{A}Z^{B}\mathcal{M}^{-2}\right),\nn\\
{\bf 10}_{-2}:&&\quad
  {\rm Tr}\left(Z^{\dagger}_A Z^{\dagger}_B\mathcal{M}^{2}\right),
\eea
where $A$ and $B$ label the ${\bf 4}$ of $SU(4)$, while
$\mathcal{M}^{-2}$ and $\mathcal{M}^{2}$ are the monopole operators
in the proper representations of the gauge group needed for
gauge invariance of
the operators \cite{Klebanov:2009sg}. When the gauge group is
$SU(2)\times SU(2)$, the bifundamental scalars of the ABJM theory are
related to the original BLG variables with $SO(4)$ indices through
\be
Z^A=\phi^A+\im \phi^{A+4},\qquad
\phi^I=\ft12(\phi^I_4\oneone+\im \phi_i^I\sigma^i),
\ee
where the $SO(8)$ ${\bf 8}_v$ indices are raised using $\delta^{IJ}$.
Similarly, one can reformulate the ${\bf35}_s$ operators of $SO(8)$ in
the framework of ABJM.

\subsection{Uplift to eleven dimensions }

The ${\bf 35}_v$ dilatons parameterise the coset $SL(8,\mathbf{R})/SO(8)$, and
one can use the local $SO(8)$ symmetry to diagonalise the coset, so that
the scalar Lagrangian can be written in terms of seven dilatons
$\vec{\varphi}$
\be\label{L2}
\mathcal{L}=R-\frac{1}{2}(\partial\vec{\varphi})^2-V,
\ee
where
\be
V=-\frac{1}{8}g^2\left((\sum_{i=1}^8 X_i)^2-2\sum_{i=1}^8 X_i^2\right).
\ee
The eight $X_i$, subject to the constraint
\be
\prod_{i=1}^8 X_i=1,
\ee
are parameterized by the seven dilatons as
\be
X_i=e^{-\frac{1}{2}\vec{b}_i \cdot\vec{\varphi}},
\label{cosetpara}
\ee
where $\vec{b}_i$ are the weight vectors of the fundamental
representation of $SL(8,\mathbf{R})$, satisfying
\be
\vec{b}_i\cdot \vec{b}_j=8\delta_{ij}-1,
\qquad\sum_i\vec{b}_i=0,\qquad \sum_i(\vec{u}\cdot\vec{b}_i)\,
   \vec{b}_i=8\vec{u},
\ee
and $\vec{u}$ is an arbitrary vector. Setting
\be
X_1=X_2=X_3=X_4=e^{\frac{1}{2}\phi_1},\quad X_5=X_6=X_7=e^{-\frac{1}{2}(\phi_1-2\phi_2)},\quad X_8=e^{-\frac{1}{2}(\phi_1+6\phi_2)},
\ee
the Lagrangian \eqref{L2} becomes equivalent to our Lagrangian \eqref{L1}
in the case that $\sigma_1=\sigma_2=0$.

In \cite{Cvetic:1999xx}, a set of domain-wall solutions was found,
given by
\bea
ds^2&=&(\ft12gr)^4(\prod_i H_i)^{1/4}dx^{\mu}dx_{\mu}+(\prod_i H_i)^{-1/4}\frac{4dr^2}{g^2r^2},\nn\\
X_i&=&H_i^{-1}(\prod_{j=1}^{8}H_j)^{1/8},
\label{domainwallsln}
\eea
where
\be
H_i=1+\frac{\ell_i^2}{r^2}\nn.
\ee
The transformation
\be
r^2\rightarrow r^2+\eta,\quad \ell_i^2\rightarrow \ell_i^2-\eta,
\label{shifttrans}
\ee
is a diffeomorphism, and so the inequivalent solutions are parameterized
by only seven, rather than eight, of the constants $\ell_i^2$.  In
particular, $\eta$ can be chosen so that the smallest of the $\ell_i^2$ is
set to zero, while keeping the remaining ones non-negative.

    The domain-wall solutions are singular in the IR, and the nature of
the singularity depends on the number of $\ell_i^2$ that can be set to zero
by means of the shift symmetry (\ref{shifttrans}). In terms of the new
coordinates, where the smallest of the $\ell_i^2$ have been shifted to zero,
the singular IR behavior of the metric is given by
\be
ds_4^2\simeq R^{\ft{8-k}2}dx^{\mu}dx_{\mu}+R^{\ft{k-4}2}dR^2=\rho^{\ft{2(8-k)}k}dx^{\mu}dx_{\mu}+d\rho^2,
\ee
where $k$ is the number of $\ell_i^2$ that are set to zero by the shift.
To compare with (\ref{boundaryseries}), we expand
$S\equiv\tanh\frac{1}{2}\phi_1$ and
$\widetilde{S}\equiv\tanh\frac{1}{2}\phi_2$ in terms of $z$, which is
related to $r$ by
\be
\frac{dr}{dz}=-\frac{r}{2z}(\prod_i H_i)^{1/8}.
\ee
We find that
\bea
S(z)&=&=\ft18(-4a+3b+c)z+\ft1{16}(-28a^2+15b^2+12bc+c^2)z^2+\cdots , \nn\\
\widetilde{S}(z)&=&\ft18(c-b)z+\ft1{16}(-8ab-5b^2+8ac+4bc+c^2)z^2+\cdots , \nn
\label{boundaryseries2}
\eea
where
\be
\ell_1^2=\ell_2^2=\ell_3^2=\ell_4^2=a,\qquad
\ell_5^2=\ell_6^2=\ell_7^2=b\,,\qquad \ell_8^2=c.
\label{lsquared}
\ee
By setting
\be
a=-\alpha_1,\quad b=\alpha_1-2\tilde{\alpha}_1,\quad c=\alpha_1+6\tilde{\alpha}_1,
\label{lconst}
\ee
we see that (\ref{lconst}) reproduces (\ref{boundaryseries}).
From (\ref{boundaryseries2}), one can see that the leading coefficients are
invariant under the shift (\ref{shifttrans}), and that they therefore
have an invariant physical meaning. Below, we shall show that in fact
they are related to the VEVs of the ${\bf 35}_v$ operators.

 The solution in (\ref{domainwallsln}) can be uplifted to eleven dimensions,
where it describes a continuous distribution of M2-branes
\cite{Cvetic:1999xx}. The uplifted solution is given by
\be
d\hat{s}_{11}^2=H^{-2/3}(-dt^2+d\vec{x}\cdot d\vec{x})+H^{1/3}ds_8^2,\qquad\hat{F}=dt\wedge dx^1\wedge dx^2\wedge dH^{-1},
\label{liftmetric}
\ee
where
\be
H=\frac1{(g/2)^6\, r^6\, \Delta}\qquad
\Delta=(H_1\cdots H_8)^{1/2}\, \sum_{i=1}^8 \frac{\mu_i^2}{H_i}.
\ee
The transverse-space metric $ds^2_8$ is given by
\be
ds^2_8=\frac{\Delta \,dr^2}{\sqrt{H_1\cdots H_8}}+r^2\,
    \sum_{i=1}^8 H_i d\mu_i^2,\label{tranmet}
\ee
where $\sum_i \mu_i^2=1$ defines a unit $S^7$ in $\R^8$.
The metric (\ref{tranmet})  can be expressed as a flat Euclidean
8-metric $ds^2_8=dy^mdy^m$ by making the coordinate transformation
\be
y_i=r\sqrt{H_i}\, \mu_i.
\label{r2y}
\ee
In terms of these Euclidean coordinates, the harmonic function $H$
takes the form
\be
H=(g/2)^{-6}\, \int\frac{\sigma (\vec{y}\, ')d^8y'}{|\vec{y}-\vec{y}\,'|^6},
\ee
where $\sigma$ is the normalized distribution function of the M2-branes.

    Besides the trivial coincident branes case described by a delta-function
distribution, solutions in our truncated theory correspond to
six possible distributions of M2-branes, depending on the relative
magnitudes of the constants $\ell_i^2$, which are given in (\ref{lsquared}).

\begin{itemize}

\item $\ell^2_8$ is the smallest among the $\ell^2_i$. In this case,
using the shift symmetry (\ref{shifttrans}), $\ell_8^2$ can be set to zero,
and the M2-branes are distributed in a 7-ellipsoid. The explicit form of
the distribution function can be found in \cite{Cvetic:1999xx}, which
suggests the existence of branes with negative tension. Thus, solutions
in this class are unphysical.

\item $\ell_5^2=\ell_6^2=\ell_7^2$ are the smallest. In this case,
$\ell_5^2=\ell_6^2=\ell_7^2$ can be set to zero by the shift symmetry,
and the resulting geometry describes M2-branes distributed in a 5-ellipsoid.
The distribution function is positive definite.

\item  $\ell_1^2=\ell_2^2=\ell_3^2=\ell_4^2$ are the smallest. In this case,
$\ell_1^2=\ell_2^2=\ell_3^2=\ell_4^2$ can be set to zero by shift symmetry.
Accordingly, we find M2-branes with positive tension, distributed in
a 4-ellipsoid.

\item  $\ell_5^2=\ell_6^2=\ell_7^2=\ell_8^2$ are the smallest. In fact,
this case corresponds to the $SO(4)$ truncation of ${\cal N}=8$
supergravity \cite{Cvetic:1999au}. The domain-wall solution is sourced
by M2-branes distributed in a 4-ball with positive tension.
  \item   $\ell_1^2=\ell_2^2=\ell_3^2=\ell_4^2=\ell_8^2$ are the smallest. This case describes positive-tension M2-branes distributed in a 3-ball.

  \item   $\ell_1^2=\ell_2^2=\ell_3^2=\ell_4^2=\ell_5^2=\ell_6^2=\ell_7^2$
are the smallest. This case is contained in the $G_2$ truncation of
${\cal N}=8$ supergravity \cite{Ahn:2001kw}.  The domain-wall solution
is sourced by positive-tension M2-branes distributed in a segment.
\end{itemize}

To summarize, except for the case where the M2-branes are distributed
in a 7-ellipsoid, the domain-wall solutions are all sourced by M2-branes
with positive tension, and therefore they may be considered to be physical.

 The harmonic function $H$ has a Taylor expansion at large $|\vec{y}\, |$
given by
\be
H=\frac{1}{(g/2)^6|\vec{y}\,|^6}
\Big(1+\sum_{n=2}^{\infty} \frac{2^n(n+1)(n+2)d^{(n)}_{i_1i_2\dots i_n}
\hat{y}^{i_1}\hat{y}^{i_2}\dots\hat{y}^{i_n}}{|\vec{y}\,|^{n}}\Big),
\ee
where $\hat{y}^i\equiv y^i/|\vec{y}\,|$, and the partial-wave expansion
coefficients $d^{(n)}_{i_1i_2\dots i_n}$ are totally symmetric and traceless,
transforming in the $(n,0,0,0)$ representation under $SO(8)$.
The scaling dimension assigned to $d^{(n)}_{i_1i_2\dotsi_n}$ is $\Delta=n/2$
rather than $\Delta=n$, since at large $\vec{y}$, $|\vec{y}\,|^2$
is asymptotic to the standard radial coordinate of AdS. In general,
the higher-order coefficients $d^{(n)}_{i_1\cdots i_n}$ with $n>2$
are present, which means there are VEVs for higher-dimension operators as
well as for those in the ${\bf 35}_v$. Therefore, although the
consistent truncation keeps only a finite number of fields, the profile
of the Coulomb branch flow captures infinitely many VEVs.
For our case, from (\ref{r2y}), we find
\be
r=|\vec{y}\,|\Big(1-\frac{a(\hat{y}_1^2+\hat{y}^2_2+\hat{y}_3^2+\hat{y}_4^2)
 +b(\hat{y}_5^2+\hat{y}^2_6+\hat{y}_7^2)+c\hat{y}_8^2}{2|\vec{y}\,|^2}
    +\cdots\Big).
\ee
Using the equation above, we obtain the partial-wave expansion of the
harmonic function $H$ up to ${\cal O}(1/|\vec{y}\,|^8)$,  in which
the coefficients $d^{(2)}_{i_1i_2}$ are given by
\bea
d_{i_1i_2}^{(2)}&=&\ft1{48}\,
 {\rm diag}\Big((4a-3b-c)\times\oneone_{4},(5b-4a-c)\times
     \oneone_3, 7c-4a-3b\Big),\nn\\
&=&\ft16 \,{\rm diag}\Big(-\alpha_1\times\oneone_{4},
   (\alpha_1-2\widetilde{\alpha}_1)\times\oneone_3,
  -(\alpha_1+6\widetilde{\alpha}_1)\Big).
\eea

\subsection{Holographic RG flow with ${\cal N}=1$ mass deformations}

    In this section, we study the solutions to eq.~(\ref{dweq}) when the
pseudoscalars are turned on. In this case, besides the non-trivial
AdS$_4$  $G_2$ critical point found in \cite{Warner:1983vz}, the other
interesting solutions are domain walls asymptotic to AdS$_4$ in the
ultraviolet. Specifically, for the domain-wall solutions we shall
consider\footnote{Domain-wall solutions interpolating between two
supersymmetric AdS vacua have been studied in \cite{Ahn:2001kw, Bobev:2009ms}.},
the pseudoscalars $P$ and $\widetilde{P}$ behave
like $P(z)\sim \beta_1 z$ and $\widetilde{P}\sim \widetilde{\beta}_1 z$ near
the AdS$_4$ boundary.
These solutions are dual to RG flows on M2-branes driven by ${\cal N}=1$
mass deformations. An exact solution of this type was found in
\cite{Pope:2003jp}, in the $SO(4)$ gauged ${\cal N}=4$ supergravity.
As we shall see, domain-wall solutions with nontrivial pseudoscalar
profiles are generically singular at a certain IR
cutoff $z_{\rm IR}$. It turns out that the metrics of
these solutions share similar singular behaviors with
the 4D Coulomb branch flow metrics.

   In general, the IR cutoff is attained when the moduli of the two
complex scalars $\psi_1$ and $\psi_2$ defined in (\ref{psidef})
approach infinity. To visualize these flows, we associate
each complex scalar with a complex plane. It turns out that the degree of
the IR singularity depends on the direction in which the complex scalar
approaches infinity in the complex plane. There are also some special
flows in which only one of the complex scalar fields blows up at
the IR cutoff, while the other field tends to a finite value.
Owing to the complexity of the flow equations (\ref{dweq}), we are not able
to classify all possible solutions. Instead, we shall present several
representative
examples of solutions that exhibit different types of singular IR behaviors.

We shall use the coordinates defined in (\ref{dwm}), in which the
IR cutoff $z_{\rm IR}$ corresponds to $\rho=\rho_{\rm IR}$.
Under the diffeomorphism $\rho\rightarrow\rho+\lambda$,  the IR
cutoff $\rho_{\rm IR}$ and Fefferman-Graham expansion coefficients
are changed to
\be
\rho_{\rm IR}\rightarrow \rho_{\rm IR}-\lambda,\qquad (\alpha_1,\beta_1,
\widetilde{\alpha}_1, \widetilde{\beta}_1)\rightarrow
e^{-\lambda}\, (\alpha_1,\beta_1, \widetilde{\alpha}_1, \widetilde{\beta}_1).
\ee
Therefore, the combinations
$e^{-\rho_{\rm IR}}(\alpha_1,\beta_1, \widetilde{\alpha}_1,
\widetilde{\beta}_1)$ are invariant under the shift of $\rho$, and they
characterise the solution. By shifting $\rho$ one can choose a special
coordinate in which $\rho_{\rm IR}=0$, and the Fefferman-Graham
coefficients $(\alpha_1,\beta_1, \widetilde{\alpha}_1, \widetilde{\beta}_1)$
are then equal to the shift-invariant quantities. We shall present
the domain-wall solutions using this particular choice of coordinate.
For technical convenience, we work with the redefined the scalar fields
\be
\zeta_1=\tanh\ft12\phi_1 \, e^{-\im \sigma_1},\quad
\zeta_2=\tanh\ft12\phi_2 \, e^{\im \sigma_2}.
\ee
Infinity in the complex $\psi$ plane is mapped into the unit circle in
the complex $\zeta$ plane.

\begin{figure}[h]
\centering \includegraphics[width=8cm]{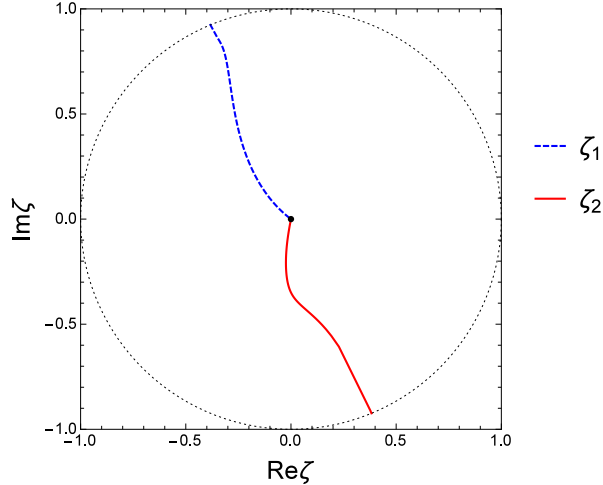}
\caption{\it An example of the most common flow with both $\phi_1$ and $\phi_2$
blowing up in the IR. This flow is obtained by choosing
$\sigma_{1{\rm IR}}=\ft{11}{8}\, \pi$ and $\sigma_{2{\rm IR}}=\ft{13}{8}\pi$.}
\label{flow7L}
\end{figure}

   We first discuss the solutions in which $|\zeta_1|$ and $|\zeta_2|$
both approach 1 in the IR at $\rho=0$. Near the unit circle, we may
perform a perturbative expansion for $\zeta_1$ and  $\zeta_2$, writing
\be
\zeta_1=(1-\delta r_1)e^{-i(\sigma_1+\delta \sigma_1)},\qquad  \zeta_2=(1-\delta r_2)e^{i(\sigma_2+\delta \sigma_2)}.
\ee
 When  $\zeta_1$ and $\zeta_2$ approach the unit circle from generic angles
specified by $\sigma_{1{\rm IR}}$ and $\sigma_{2{\rm IR}}$, the perturbations
at leading order satisfy the equations
\bea
\frac{d(\delta r_1)}{d\rho}&=& f_1(\sigma_{1{\rm IR}},\sigma_{2{\rm IR}}) \frac{\delta r_1^{1/2}}{\delta r_2^3}+\cdots,\quad \frac{d(\delta r_2)}{d\rho}= f_1(\sigma_{1{\rm IR}},\sigma_{2{\rm IR}}) \frac{1}{\delta r_1^{1/2}\delta r_2^2}+\cdots, \nonumber\\
\frac{d(\delta \sigma_1)}{d\rho}&=&- f_2(\sigma_{1{\rm IR}},\sigma_{2{\rm IR}}) \frac{\delta r_1^{3/2}}{\delta r_2^3}+\cdots,\quad \frac{d(\delta \sigma_2)}{d\rho}= f_3(\sigma_{1{\rm IR}},\sigma_{2{\rm IR}}) \frac{1}{\delta r_1^{1/2}\delta r_2}+\cdots,
\eea
where the non-vanishing coefficients $f_1$, $f_2$ and $f_3$ are given by
\bea
&f_1(\sigma_1,\sigma_2)&=\sqrt{2}\sin ^2\frac{\sigma _2}{2} |f_0|,\quad  f_0=3 \sin\frac{\sigma _1}{2}+2 \sin(\frac{\sigma _1}{2}-\sigma _2)+\sin (\frac{\sigma _1}{2}-2 \sigma _2), \nonumber\\
&f_2(\sigma_1,\sigma_2)&= {\rm sign}(f_0)\sqrt{2} \sin ^2\frac{\sigma _2}{2} \Big(3 \cos \frac{\sigma _1}{2}+2 \cos (\frac{\sigma _1}{2}-\sigma _2)+\cos (\frac{\sigma _1}{2}-2 \sigma _2)\Big), \nonumber\\
&f_3(\sigma_1,\sigma_2)&={\rm sign}(f_0)\sqrt{2} \sin\frac{\sigma _2}{2}\sin (\frac{\sigma _1}{2}-\sigma _2) \Big(\cos \frac{\sigma _2}{2}+\cos \frac{3 \sigma _2}{2}\Big).
\eea
An example of a solution in this class is given in Fig. \ref{flow7L},
where $\sigma_{1{\rm IR}}=\ft{11}{8}\pi$ and
$\sigma_{2{\rm IR}}=\ft{13}{8} \pi$.
 Near the IR region $\rho=0$, solutions for the perturbations
take the form
\bea
\delta r_1&=& 0.25 \, \rho ^{2/7},\qquad \delta r_2 = 1.508507 \,
\rho ^{2/7},\nn\\
\delta \sigma_1&=& 0.106352\, \rho ^{4/7},\qquad
\delta \sigma_2 =  -0.226322 \, \rho ^{4/7},
\eea
from which we can obtain the IR behavior of the scale factor
$e^{2A(\rho)}$, and therefore the IR behavior of the metric.
It turns out that the metric shares a similar singular behavior
with the 4D
Coulomb branch flow metric whose 11D uplift describes a
continuous distribution of M2-branes on a 7-ellipsoid, namely,
\be
 \rho\rightarrow 0, \quad ds_4^2\simeq d\rho^2+ \rho^{2/7}dx^{\mu}dx_{\mu}.
\label{singularflow7L}
\ee
 By integrating the equation from the IR to the UV, we can read off the
Fefferman-Graham coefficients, finding
\be
\{\alpha_1,\beta_1,\tilde{\alpha}_1,\tilde{\beta}_1\}=\{-1.282820, 0.971868, -0.191382, -0.959326\}.
\ee
As mentioned before, since we are using a radial coordinates where
$\rho_{\rm IR}=0$, the Fefferman-Graham coefficients are equal to
the $\rho$-shift-invariant combinations.

If $\zeta_1$ and $\zeta_2$ approach the unit circle in some special
directions such that $f_1(\sigma_{1{\rm IR}}, \sigma_{2{\rm IR}})=0$,
the singular behavior of the solution
shows different features. Having
$f_1(\sigma_{1{\rm IR}}, \sigma_{2{\rm IR}})=0$ implies either
$\sigma_{2{\rm IR}}=0$ or $f_0=0$. In the former case, the equations of
motion of the perturbations near $\rho=0$ are given by
\bea
\frac{d(\delta r_1)}{d\rho}&=&\frac{3\sqrt{2} | \sin(\sigma_{1{\rm IR}}/2)|
\delta r_1^{1/2}}{2\delta r_2}+\cdots,\quad \frac{d(\delta r_2)}{d\rho}=
\frac{\sqrt{2}|\sin(\sigma_{1{\rm IR}}/2)|}{2\delta r_1^{1/2}}+\dots,
\\
\frac{d(\delta \sigma_1)}{d\rho}&=&-\frac{3 \sqrt{2}
|\cos(\sigma_{1{\rm IR}}/2)| \delta r_1^{3/2}}{2 \delta r_2}+
\cdots,\quad \frac{d(\delta \sigma_2)}{d\rho}=\frac{\sqrt{2}
|\cos(\sigma_{1{\rm IR}}/2)|\delta r_2}{3\delta r_1^{1/2}}+\cdots.\nonumber
\eea

\begin{figure}[h]
\centering \includegraphics[width=8cm]{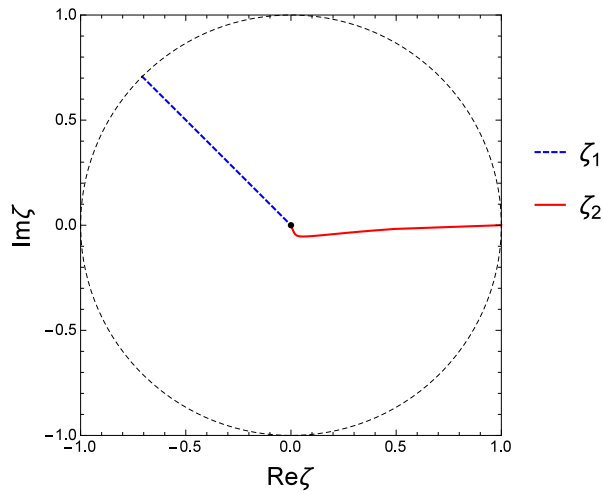}
\caption{\it A typical flow obtained by choosing
$\sigma_{1{\rm IR}}=5\pi/4$  and $ \sigma_{2{\rm IR}}=0$. }
\label{flow5L}
\end{figure}

An explicit example of a solution for the $\sigma_{1{\rm IR}}=5\pi/4$
 case is given in Fig. \ref{flow5L}.
Near the IR cutoff, the perturbations take the form
\bea
\delta r_1&=& 0.01 \, \rho ^{6/5},\qquad
\delta r_2 = 16.3320 \, \rho ^{2/5}, \nonumber\\
\delta \sigma_1&=& 2.07107\times 10^{-5}\, \rho ^{12/5},\qquad
\delta \sigma_2=-88.9118 \, \rho ^{4/5} ,
\eea
which we leads to the following singular behavior of the metric
near the IR cutoff:
\be
 ds_4^2\simeq d\rho^2+ \rho^{6/5}dx^{\mu}dx_{\mu}.
\label{singularflow5L}
\ee
This IR singularity is similar to the one appearing in the 4D
Coulomb-branch flow metric whose 11D uplift describes a
continuous distribution of
M2-branes on a 5-ellipsoid.
In this example, the Fefferman-Graham expansion coefficients of the
two complex scalars are given by
\be
\{\alpha_1,\beta_1,\tilde{\alpha}_1,\tilde{\beta}_1\}=\{-1.419990, 1.418240, 0.049763, -0.139465\}.
\ee

\begin{figure}[h]
\centering \includegraphics[width=8cm]{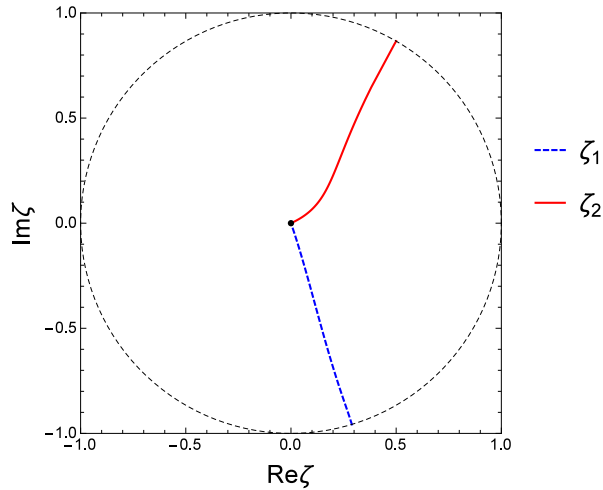}
\caption{\it A typical flow in $f_0=0$ case  given by $\sigma_{1{\rm IR}}=1.27712$, $\sigma_{2{\rm IR}}=\pi/3$.  }
\label{flow0L}
\end{figure}

An example for the $f_0=0$ case  is given by $\sigma_{1{\rm IR}}=1.27712$
and $\sigma_{2{\rm IR}}=\pi/3$. The numerical solutions for the
two complex scalars are exhibited in Fig. \ref{flow0L}.
In this solution, the IR expansion of the perturbations takes the form
\bea
\delta r_1&=&  1.0\, \rho ^{6/11},\qquad
\delta r_2= 1.15686 \, \rho ^{4/11}, \nonumber\\
\delta \sigma_1&=& 0.545275 \, \rho ^{12/11},\qquad
\delta \sigma_2= -0.193171 \, \rho ^{12/11},
\eea
which leads to a novel singular behavior of the metric which
has not been observed in the Coulomb-branch flow metric, with
\be
ds^2_4\simeq d\rho^2+\rho^{6/11} dx^{\mu}dx_{\mu}.
\label{sinularflow0L}
\ee
The UV expansion is characterised by the coefficients
\be
\{\alpha_1,\beta_1,\tilde{\alpha}_1,\tilde{\beta}_1\}=\{0.487525, -1.402800, 0.564112, 0.257166\}.
\ee

\begin{figure}[h]
\centering \includegraphics[width=8cm]{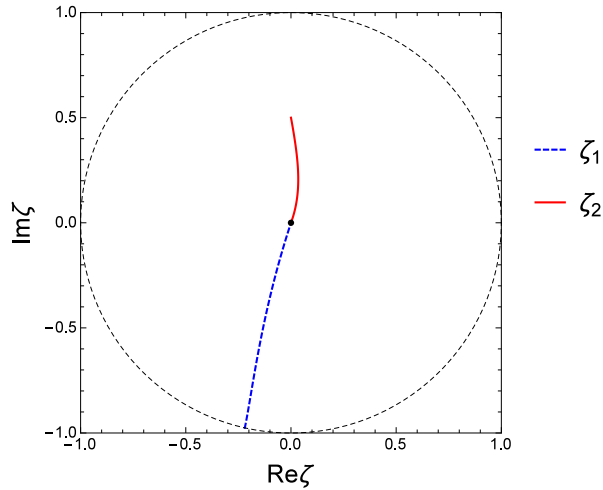}
\caption{\it A typical flow with $\zeta_{2{\rm IR}}=0.5$,
$\sigma_{1{\rm IR}}=1.792110$ and $\sigma_{2{\rm IR}}=\pi/2$.  Unlike
previous examples, $\phi_2$ remains finite in the IR limit.}
\label{flow4L}
\end{figure}

Having shown several examples in which both $\zeta_1$ and $\zeta_2$
approach the unit circle in the IR, we now present an example in which
$\zeta_2$ limits to a point inside the unit circle; in other words,
$\phi_2$ stays finite at the IR cutoff. In general, this imposes a
very complicated functional relation among $|\zeta_2|$, $\sigma_{1{\rm IR}}$
and $\sigma_{2{\rm IR}}$. However, we find that if $\sigma_{2{\rm IR}}=\pi/2$,
this expression reduces to the rather simple form
\be
\sigma_{1{\rm IR}}=2\arctan\Big(\frac{1+|\zeta_{2{\rm IR}}|^2}{2|\zeta_{2{\rm IR}}|}\Big).
\ee
For $|\zeta_{2{\rm IR}}|=0.5$, the numerical solution is given in
Fig. \ref{flow4L}.
The IR expansion of the solution takes the form
\bea
\delta r_1&=& 0.846883 \, \rho ^2,\qquad
\delta \sigma_1=-1.02003 \, \rho ^4,\nn\\
\delta {\rm Re}\zeta_2&=& 0.39589\, \rho^{1.47386},\qquad
\delta {\rm Im}\zeta_2=-1.96043 \, \rho ^{1.47386},
\eea
for which the singular IR behavior of the metric takes the form
\be
ds^2_4\simeq d\rho^2+\rho^{2}dx^{\mu}dx_{\mu}.
\label{sinularflow4L}
\ee
In this case, it should be noted that the convenient variables to
study the perturbation of $\zeta_2$ are
$(\delta {\rm Re}\zeta_2,\, \delta {\rm Im}\zeta_2)$ rather than
$(\delta r_2,\,\delta \sigma_2).$
This IR singularity is similar to the one appearing in the 4D
Coulomb-branch flow metric whose 11D uplift describes a continuous
distribution
of M2-branes on a 4-ellipsoid.
The UV expansion is specified by the coefficients
\be
\{\alpha_1,\beta_1,\tilde{\alpha}_1,\tilde{\beta}_1\}=\{-0.453355, -1.947710, 0.210519, 0.285747\}.
\ee

Now we turn to the higher-dimensional interpretation of the solutions
with non-trivial pseudoscalars, which correspond to turning on mass
deformations on M2 branes.  Pseudoscalars in the 4D supergravity appear
both in the 11D metric and the 3-form potential. Non-vanishing pseudoscalars
generally lead to non-trivial internal components of the 3-form potential.
Thus it is conceivable that the uplift of the lower-dimensional solution
with non-vanishing pseudoscalars implies the presence of M5-branes in
the bulk geometry. Previous studies
\cite{Pope:2003jp,Bena:2004jw,Lin:2004nb } provide evidence for this
intuition, with M2-branes growing into dielectric M5-branes
wrapped on $S^3$ via the Myers effect \cite{Myers:1999ps}.

When pseudoscalars are turned on, we have found four classes of solutions
exhibiting different IR singularities. In order to see whether these
singular solutions are physically allowable as brane configurations,
we shall embed our solutions into 11D supergravity. The metric ansatz
for lifting ${\cal N}=8$ gauged supergravity to
$D=11$ is given in \cite{deWit:1984nz}, with (in the absence of the
four-dimensional Yang-Mills fields)
\be
d\hat s_{11}^2 = \Delta^{-1}\, ds_4^2 + g_{mn}(x,y)\, dy^m dy^n ,
\ee
where the inverse of the internal seven-dimensional metric is given by
\be
\Delta^{-1} g^{mn}(x,y) = \ft14 \mathring{K}^{m\, IJ}\, \mathring{K}^{n\, KL}\,
  (u_{ij}{}^{IJ} + v_{ij IJ})(u^{ij}{}_{KL} + v^{ij KL}).
\label{metans}
\ee
In the above formula, the warp factor is given by
\be
\Delta=\sqrt{ \fft{\det g_{mn}(x,y)}{\det\mathring{g}_{mn}(y)}},
\ee
where $\mathring{g}^{mn}(y)$ is the metric on the unit 7-sphere and
the 28 Killing vectors $\mathring{K}^{m\, IJ}$ are those of the
unit 7-sphere. These may be described in terms of the coordinates $X^A$
on an 8-dimensional
Euclidean space, subject to the constraint $X^A X^A=1$, as
\be
\mathring{K}^{IJ} = \ft12 (\Sigma_{IJ})_{AB}\, \Big( X^A\, \fft{\del}{\del X^B}- X^B\, \fft{\del}{\del X^A}\Big).
\ee
Here $\Sigma_{IJ}$ are the $8\times 8$ chiral projection of the
$SO(8)$ 2-Gamma matrices $\Gamma_{IJ}$.  ($AB$ are spinor indices.)
The explicit form of $\Gamma_{IJ}$ and the adapted coordinates $X^A$
can be found in Appendix {\ref{conventions}}.

Using the above formulae,
we have computed the uplifted metric for the $SO(3)_D\times SO(3)_R$ invariant
sector of $SO(8)$ gauged ${\cal N}=8$ gauged supergravity. The resulting
expressions for each component of the 11D metric are long, and seem to
be unsuitable for presentation. In spite of the complexity of the 11D metric,
we can still extract its behavior near the location of the
singularity, say at $\rho=0$, using numerical techniques. This
encodes information about the brane configuration. Plugging
in specific values for the angles on the deformed $S^7$, we can read
off the singular behavior of the 11D metric. The robustness of this
singular behavior can be checked by varying the values of the angles.
By this means, we find that the uplift of the solution given in
Fig. \ref{flow7L} can be brought to the form, near $\rho=0$,
\be
d\hat s^2_{11}\simeq H^{-2/3}ds^2_{3\parallel}+H^{1/3}ds^2_{8\perp},
\qquad H=r,\quad r=\rho^{7/4}.
\ee
 This singular behavior is the same as that appearing in the metric
describing M2-branes distributed in a 7-ellipsoid, which involves
M2-branes with negative tension. Thus, the class of flow solutions
represented by the example in Fig. \ref{flow7L} is not physically
acceptable.

 For M5-branes distributed in an $n$-ellipsoid, near the IR singularity
the metric behaves like
\be
d\hat s^2_{11}\simeq H^{-1/3}ds^2_{6\parallel}+H^{2/3}ds^2_{5\perp},\quad H=r^{n-3}.
\label{M5geometry}
\ee
We find that after being uplifted to 11D, the metrics corresponding to the
solutions given in Fig. \ref{flow5L} and Fig. \ref{flow4L} can be brought
to the above forms with $n=2$ and $n=1$ respectively, by identifying
$\rho$ as the appropriate power of $r$. The six directions on the world-volume
of the M5-branes consist of three directions on the world-volume of
the M2-branes,
and three directions from the 7-sphere. Since there are only three
flat directions in the world-volume of the M5-branes, they must wrap
on the remaining three directions.   We are comparing such a
configuration of M5-branes with one in which the M5-branes
are flat and infinitely large.  This comparison is valid, as long as we
focus on the region infinitesimally close to brane. From the results
given in \cite{Cvetic:1999xx}, it can be deduced that for $n\leq3$,
the geometry (\ref{M5geometry}) is sourced by M5-branes with positive
tension. Thus solutions possessing similar singular behaviors to
the ones shown in Fig. \ref{flow5L} and Fig. \ref{flow4L} are
physically allowed, with the singularity being balanced by normal
positive-tension brane sources.

We also computed the uplift for the class of solutions represented by
Fig. \ref{flow0L}, for which the warp factor is given by
\be
\Delta\sim \rho^{-10/11}.
\ee
The 11D metric for this solution does not approximate any Coulomb-branch
metric of M-branes given in \cite{Cvetic:1999xx} in the near horizon
region,  indicating that a new supersymmetric solution should exist
in 11D supergravity.

\section{Holographic ${\cal N}=1$ RG Flows in the $\omega$-Deformed Theory }

In the $\omega$-deformed theory, it is straightforward to verify
that in the second-order equations of motion, $ \sigma_1$ and
$ \sigma_2$ can be consistently to set zero.
However, one also can show that
\be
\frac{\partial |W|}{\partial \sigma_1}\Big|_{\sigma_1=\sigma_2=0}\neq 0,
\qquad \frac{\partial |W|}{\partial \sigma_2}\Big|_{\sigma_1=\sigma_2=0}
\neq 0,
\ee
and so there does exist a set of first-order equations which resembles the one governing the holographic
Coulomb-branch flow on M2-branes. This is
therefore a sharp distinction between the $\omega$-deformed theory and
the undeformed theory. Since we are interested in the supersymmetric
holographic RG flows, it follows that pseudoscalars will always
play a non-trivial role in the domain-wall solutions.
A supersymmetric domain-wall solution
interpolating between the $SO(8)$ point and the $G_2$ point in the
$\omega$-deformed theory has
been studied in \cite{Guarino:2013gsa}.

Owing to the lack of any known higher-dimensional origin for the
$\omega$-deformed $SO(8)$ gauged $\cN=8$ supergravity, it is unclear
which 3D CFT should be its holographic dual. However, if we assume that
such a dual CFT exists, and that it belongs to a certain type of Chern-Simons
matter theory, the bulk scalar fields should still be dual to scalar
operators that are bilinear in the boundary scalars or fermions,
as was discussed in Sect. 3 for the undeformed case.

   This assumption
leads us to a preferred redefinition of the bulk scalar fields,
for the following reason. Using the current definition of the scalar fields,
near the boundary of AdS$_4$ the functional relations amongst the
Fefferman-Graham coefficients will ostensibly depend on the $\omega$
parameter. However, as in the $\omega=0$ case,
via the AdS/CFT correspondence these functional relations should match with
those relating the bosonic mass parameters to the fermionic ones. These
are determined by the boundary $\cN=1$ supersymmetry, and should
therefore be independent of $\omega$.  This  suggests that the ostensible
$\omega$-dependence of the functional relations amongst the Fefferman-Graham
coefficients should be a technical artifact that can be removed
by redefining the bulk scalar fields. Indeed, in terms of the new
complex scalar fields
\be
\zeta_1'=e^{\ft23\im \omega}\zeta_1 ,
\qquad \zeta_2'=e^{\ft23\im \omega}\zeta_2,
\label{newzeta}
\ee
the equalities (\ref{bosonmass1}) are maintained.
As a consistency check, if we write
\be
\zeta_1'=S'+\im P',\qquad \zeta_2'=\widetilde{S}'+\im \widetilde{P}',
\ee
then in order that $S'$, $\widetilde{S}'$ and
$P'$, $\widetilde{P}'$ should be dual to
$\Delta=1$ and $\Delta=2$ primary operators respectively,
the boundary conditions
\be
S'\sim z\,  S^{(1)'},\qquad \widetilde{S}'\sim z \, \widetilde{S}^{(1)'},
\qquad P'\sim z^2 \, P^{(2)'},\qquad
\widetilde{P}'\sim z^2\, \widetilde{P}^{(2)'},
\ee
should preserve the $\cN=1$ supersymmetry of $\omega$-deformed
supergravity in the bulk. In terms of the original fields, the
above boundary conditions amount to
\bea
\cos\ft23\omega\,  S^{(2)}_1-\sin\ft23\omega \, P^{(2)}_1&=&0,
\qquad \cos\ft23\omega \, S^{(2)}_2-\sin\ft23\omega \, P^{(2)}_2=0,\nn\\
\cos\ft23\omega \, P^{(1)}_1+\sin\ft23\omega \, S^{(1)}_1&=&0,
\qquad\cos\ft23\omega \, P^{(1)}_2+\sin\ft23\omega \, S^{(1)}_2=0.
\eea
These are consistent with the $\cN=1$ boundary conditions given in
\cite{Borghese:2014gfa} for the $\omega$-deformed $SO(8)$ gauged
$\cN=8$ supergravity.

  Similarly to discussion we gave in the $\omega=0$ case,
in the study of supersymmetric domain-wall
solutions, when $\omega\ne0$ we shall now work with the
fields $\zeta_1'$ and $\zeta_2'$. Infinity in the complex
$\psi'$ plane is mapped into the unit circle in the
complex $\zeta'$ plane. From now on, for simplicity of presentation,
we shall remove the primes from
the fields; all the fields below should be interpreted as the
$\omega$-rotated ones defined in  (\ref{newzeta}). Flow solutions
in $\omega$-deformed theory are driven by the $\omega$-dependent
``potential'' $|W|$, and so the perturbative IR expansion of the
equations of motion depends on $\omega$. However, the singular
IR behaviors of the solutions are similar to those arising in
the $\omega=0$ case. For instance, for a generic solution in which
both $\zeta_1$ and $\zeta_2$ attain the unit circle, the perturbative
IR expansion is given by
\bea
\frac{d(\delta r_1)}{d\rho}&=& \tilde{f}_1(\sigma_{1{\rm IR}},\sigma_{2{\rm IR}}) \frac{\delta r_1^{1/2}}{\delta r_2^3}+\cdots,\quad \frac{d(\delta r_2)}{d\rho}= \tilde{f}_1(\sigma_{1{\rm IR}},\sigma_{2{\rm IR}}) \frac{1}{\delta r_1^{1/2}\delta r_2^2}+\cdots, \nn\\
\frac{d(\delta \sigma_1)}{d\rho}&=&- \tilde{f}_2(\sigma_{1{\rm IR}},\sigma_{2{\rm IR}}) \frac{\delta r_1^{3/2}}{\delta r_2^3}+\cdots,\quad \frac{d(\delta \sigma_2)}{d\rho}= \tilde{f}_3(\sigma_{1{\rm IR}},\sigma_{2{\rm IR}}) \frac{1}{\delta r_1^{1/2}\delta r_2}+\cdots,
\eea
where the non-vanishing $\omega$-dependent coefficients
$\tilde{f}_1$, $\tilde{f}_2$ and $\tilde{f}_3$ are given by
\bea
&\tilde{f}_1(\sigma_1,\sigma_2)&=\frac{1}{2\sqrt{2}}|\sin (\frac{\sigma _1}{2}-3 \sigma _2+\frac{4 \omega }{3})+3 \sin (\frac{\sigma _1}{2}+\sigma _2-\frac{4\omega}{3} )-4 \sin (\frac{\sigma _1}{2}+\frac{4 \omega }{3} )|,\nn\\
&\tilde{f}_2(\sigma_1,\sigma_2)&=\frac{{\rm sign}(f_0)}{2\sqrt{2}}\cos (\frac{\sigma _1}{2}-3 \sigma _2+\frac{4 \omega }{3})+3 \cos (\frac{\sigma _1}{2}+\sigma _2-\frac{4\omega }{3} )-4 \cos (\frac{\sigma _1}{2}+\frac{4 \omega}{3} ), \nn\\
&\tilde{f}_3(\sigma_1,\sigma_2)&=\frac{{\rm sign}(f_0)}{2\sqrt{2}}\sin \left(\frac{\sigma _1}{2}-\sigma _2\right) \sin (2 \sigma _2-\frac{4 \omega }{3}).
\eea

\begin{figure}[h]
\centering \includegraphics[width=8cm]{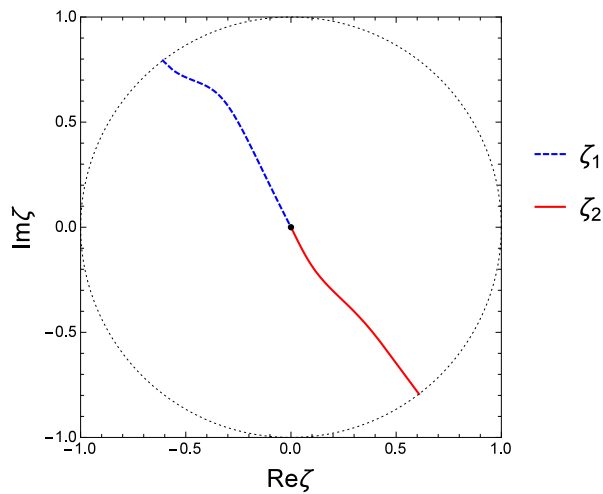}
\caption{\it A typical flow with $\sigma_{1{\rm IR}}=\frac{31 \pi }{24}$, $ \sigma_{2{\rm IR}}=\frac{41 \pi }{24}$ in the $\omega=\pi/8$ theory . }
\label{flow7Ld}
\end{figure}

\begin{figure}[h]
\centering \includegraphics[width=8cm]{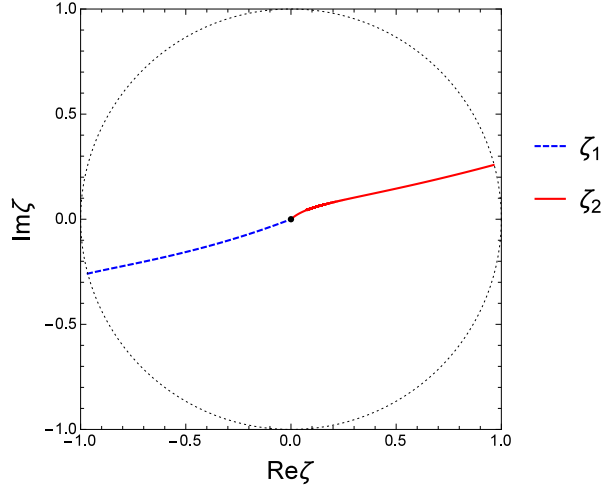}
\caption{\it A flow solution with $\sigma_{1{\rm IR}}=\pi-\frac{2\omega}{3},\sigma_{2{\rm IR}}=\frac{2\omega}{3}$ in the $\omega=\pi/8$ theory. }
\label{flow5Ld}
\end{figure}

An example of such flow solution with $\sigma_{1{\rm IR}}=31\pi/24$ and
$\sigma_{2{\rm IR}}=41 \pi/ 24$ in the $\omega=\pi/8$ theory is
exhibited in Fig. \ref{flow7Ld},
The solutions to the perturbation equation are given at the leading order by
\bea
\delta r_1=& 0.49\, \rho ^{2/7}&,\qquad
\delta r_2= 1.34846\, \rho ^{2/7}\nonumber\\
\delta \sigma_1=& 0.551838 \, \rho ^{4/7}&,\qquad
\delta \sigma_2= 0.180846 \, \rho ^{4/7}.
\eea
The metric is again of the form \eqref{singularflow7L} and the UV
expansion coefficients obtained by integrating the equations of motion
from the IR to the UV are given by
\be
\{\alpha_1,\beta_1,\tilde{\alpha}_1,\tilde{\beta}_1\}=\{-0.636249, 1.23047, 0.212564, -0.424938\}.
\ee
When $\tilde{f}_1$ vanishes, there are two branches of solutions.
The first branch is obtained when $\sigma_{1{\rm IR}}=\pi-2\omega/3$
and $\sigma_{1{\rm IR}}=2\omega/3$. For each fixed non-zero $\omega$,
this branch consists of a one-parameter family of flow solutions, modulo
the shift symmetry. However, when $\omega=0$, as we discussed in the
previous section, the solution can acquire an additional parameter and become
a two-parameter solution. In the vicinity of the IR cutoff, the
perturbations satisfy
\bea
\frac{d(\delta r_1)}{d\rho}&=&\frac{3\sqrt{2}\cos\omega\delta r_1^{1/2}}{2\delta r_2}+\cdots,\quad \frac{d(\delta r_2)}{d\rho}=\frac{\sqrt{2}\cos\omega}{2\delta r_1^{1/2}}+\dots, \nn\\
\frac{d(\delta \sigma_1)}{d\rho}&=&-\frac{2\sqrt{2}\tan\omega\delta r_1^{3/2}}{ \delta r_2^3}+\cdots,\quad \frac{d(\delta \sigma_2)}{d\rho}=\frac{2\sqrt{2}\tan\omega\delta r_2}{6\delta r_1^{1/2}}+\cdots.
\eea
For $\omega=\pi/8$, a numerical solution belonging to this branch is plotted in Fig. \eqref{flow5Ld}.
Near the IR cutoff, solutions to the perturbation equations are given as
\bea
\delta r_1=& 0.81\, \rho ^{6/5}&,\qquad \delta r_2= 1.81467\, \rho ^{2/5},\nn\\
\delta \sigma_1=& -0.0893271\, \rho ^{8/5}&,\qquad
\delta \sigma_2= 0.492134 \,\rho ^{4/5}.
\eea
This leads to an IR singularity of the metric that takes
the same form as \eqref{singularflow5L}.
Near the UV boundary, the expansion coefficients are found to be
\be
\{\alpha_1,\beta_1,\tilde{\alpha}_1,\tilde{\beta}_1\}=\{-1.18276, -0.453272, 0.19352, 0.176533\}.
\ee

\begin{figure}[h]
\centering \includegraphics[width=8cm]{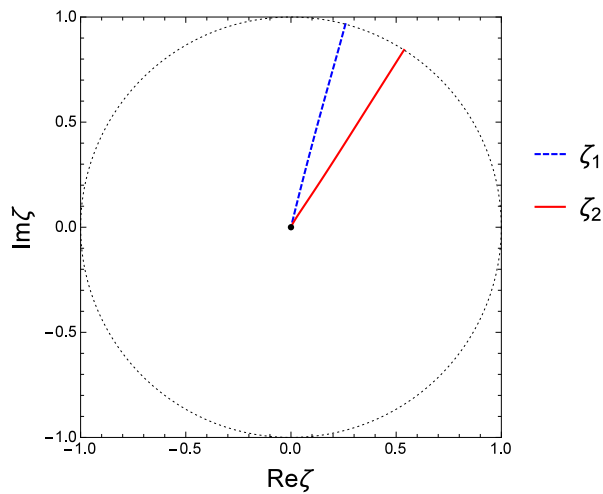}
\caption{\it A typical flow obtained in the $\omega=\pi/8$ theory by choosing $\sigma_{1{\rm IR}}=\frac{19\pi }{12}$  and $ \sigma_{2{\rm IR}}=0.319382\pi$.}
\label{flow0Ld}
\end{figure}

 When $(\sigma_{1{\rm IR}}, \sigma_{2{\rm IR}})\neq
(\pi-\frac{2\omega}{3}, \frac{2\omega}{3})$ but still with
$\tilde{f}_1$ vanishing, one obtains the second branch of solutions,
 which is analogous to those in the $\omega=0$ case with $f_0=0$
but $\sigma_{2{\rm IR}}\neq0$. A solution belonging to this branch in
the $\omega=\pi/8$ theory is found when
$\sigma_{1{\rm IR}}=\frac{19\pi }{12}$ and  $\sigma_{2{\rm IR}}=0.319382\pi$.
The numerical solution is shown in Fig. \ref{flow0Ld}
with the IR expansion
\bea
\delta r_1&=&\rho ^{6/11},\qquad \delta r_2= 1.96456 \,\rho ^{4/11}\nn\\
\delta \sigma_1&=& 0.0511178 \,\rho ^{12/11},\qquad
\delta \sigma_2= -0.0426812 \, \rho ^{12/11}.
\eea
Accordingly, near the IR singularity, the metric is of the form given in
\eqref{sinularflow0L}. The UV expansion is characterised by the coefficients
\be
\{\alpha_1,\beta_1,\tilde{\alpha}_1,\tilde{\beta}_1\}=\{0.195555, 0.801957, 0.0412681, 0.101782\}.
\ee

\begin{figure}[h]
\centering \includegraphics[width=8cm]{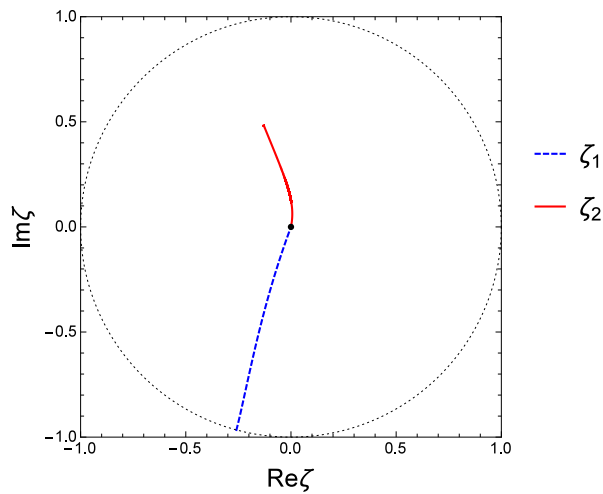}
\caption{\it A typical flow in $\omega=\pi/8$ theory obtained by choosing  $\sigma_{1{\rm IR}}=0.58351 \pi$  and $ \zeta_{2{\rm IR}}=\frac{\sqrt{2}-\sqrt{6}}{8} +{\rm i} \frac{\sqrt{2+\sqrt{3}}}{4}$. }
\label{flow4Ld}
\end{figure}

Similarly to the $\omega=0$ case, the $\omega$-deformed theory also
admits solutions in which $\zeta_2$ limits to a point inside the
unit circle; in other words
$\phi_2$ stays finite at the IR cutoff. In general, this imposes a
very complicated functional relation among $|\zeta_2|$, $\sigma_{1{\rm IR}}$
and $\sigma_{2{\rm IR}}$. However, we find that if
$\sigma_{2{\rm IR}}=\frac{2\omega}{3}$,
this expression becomes the simpler one
\be
\quad (|\zeta_{\rm IR}| ^2+1) \sin(\sigma _{1{\rm IR}}-\frac{4\omega}{3})+2 |\zeta_{\rm IR}| \cos(\sigma _{1{\rm IR}}-\frac{2\omega}{3})-2 |\zeta_{\rm IR}|  \cos 2 \omega =0.
\ee
An example of such a flow in $\omega=\pi/8$ theory is plotted in
Fig. \ref{flow4Ld},
for which the  IR behavior is given by
\bea
\delta r_1&=& 3.14822 \, \rho ^2,\qquad
\delta \sigma_1=-2.87451 \, \rho ^4\nn\\
\delta {\rm Re}\zeta_2&=& 3.90762\,  \rho ^{1.75672},\qquad
\delta {\rm Im}\zeta_2= -9.20492 \, \rho ^{1.75672}.
\eea
In this case, it should be noted that the convenient variables to study
the perturbation of $\zeta_2$ are
$(\delta {\rm Re}\zeta_2,\, \delta {\rm Im}\zeta_2)$ rather than
$(\delta r_2,\,\delta \sigma_2)$. When $\rho\rightarrow0$, the metric
appears to be of the same singular form as \eqref{sinularflow4L}.
The UV expansion of this solution is characterised  by
\be
\{\alpha_1,\beta_1,\tilde{\alpha}_1,\tilde{\beta}_1\}=\{-0.985129, -1.38687, 0.179916, 0.365543\}.
\ee

  To achieve a proper understanding of the nature of the IR singularities
of the geometry, it seems to be indispensable to embed the
lower-dimensional solution into
the UV-complete string or M-theory. Some tentative lower-dimensional
criteria for characterising a physically-allowable
IR singularity without reference to string theory were proposed in
\cite{Gubser:2000nd}. However, as pointed out by \cite{Polchinski:2010hw},
there exist solutions that satisfy these lower-dimensional criteria
but which nonetheless lift to higher-dimensional solutions with
unphysical singularities. Since the higher-dimensional origin of the
$\omega$-deformed theories is currently unknown, we must necessarily
postpone attempting to give a complete interpretation of the
$\omega$-deformed supersymmetric domain-wall solutions
which generically have IR singularities.

\section{Conclusions}

    This paper has addressed the problem of finding new solutions in the
$\omega$-deformed $SO(8)$ gauged $\cN=8$ supergravities. For the the
solutions to have holographic dual interpretations, our focus has been on
new AdS stationary points and supersymmetric domain-wall solutions
asymptotic to the $\cN=8$ AdS vacuum in the UV.  The investigation of the
new solutions in the complete $\cN=8$ supergravity is extremely complicated,
since the theory contains 70 scalars. In order to render the problem
tractable, we considered a consistently-truncated sector of the $SO(8)$
gauged $\cN=8$ supergravity, by keeping the fields invariant under the
subgroup $SO(3)_D\times SO(3)_R$ of $SO(8)$. By construction, the truncated
theory preserves $\cN=1$ supersymmetry. Besides the metric, the bosonic
sector of the truncated theory consists of four scalar fields,
parameterising an $\frac{SL(2,R)}{SO(2)}\times \frac{SL(2,R)}{SO(2)}$ coset.
The scalar potential depends on the $\omega$-parameter explicitly,
and can be reformulated using an $\omega$-dependent superpotential in the
standard way.  We gave an extensive discussion of the stationary points of
the scalar potential. In addition to the previously-known $G_2$ or
$SO(7)$-invariant points, there are two $SO(3)_D\times SO(3)_R$-invariant
critical points captured by the $SO(3)_D\times SO(3)_R$-invariant sector.
One of them preserves $\cN=3$ supersymmetry in the full $\cN=8$
theory \cite{Gallerati:2014xra}, while the other one, which had not been
found previously, is non-supersymmetric but nonetheless stable. The
cosmological constants of these two critical points depend on
the value of the $\omega$ parameter.  In each case the value of
the cosmological constant diverges in the $\omega\rightarrow 0$ limit,
indicating the absence of these two new stationary points in the original
de Wit-Nicolai theory.

    We then looked for supersymmetric domain-wall solutions, which satisfy
a set of first-order equations required by the $\cN=1$ supersymmetry.
These equations describe a gradient flow in the scalar coset manifold,
with the superpotential being the ``potential'' whose gradient drives the
flow. A general feature of the domain-wall solutions is that the
scalar fields flow to infinite values at an IR cutoff where the metric
becomes singular.

    In the original de Wit-Nicolai theory, supersymmetry of the flow
solution does not force the pseudoscalar fields to be turned on. In fact,
when the pseudoscalars are turned off, explicit supersymmetric domain-wall
solutions exist, and they preserve enhanced 16 supercharges.
These solutions can lift to solutions in M-theory describing the
Coulomb-branch flows on M2-branes spreading out into six possible
distributions in the transverse space \cite{Cvetic:1999xx}. We showed
explicitly that the VEVs of the dimension-1 primary operators driving the
Coulomb-branch flow are encoded in the leading UV expansion coefficients
of the scalar fields of the 4-dimensional theory. The physically-allowed
IR-singular solutions are the ones sourced by M2-branes of positive tension.

   When the pseudoscalars are turned on, the complexity of the flow
equations prevents us from obtaining the solutions analytically. We were
only able to obtain numerical solutions, by integrating the flow equations
from the IR to the UV. The solutions we found correspond to RG flows driven
by both the VEVs and also by supersymmetric mass deformations. The
supersymmetric mass terms correspond to the non-vanishing UV expansion
coefficients of the pseudoscalars. The competition between the VEVs
and the mass terms leads to a variety of possible IR singularities in the
geometry. Lifting to M-theory, the physical solutions approach
the Coulomb-branch flow of dielectric M5-branes wrapping on $S^3$ in the deep IR.

In the $\omega$-deformed theories, within the same truncated scalar sector,
supersymmetry of the domain-wall solutions requires that the pseudoscalars
be active.
The singular IR behaviors of the solutions are similar to those arising
in the
$\omega=0$ case. However, a proper understanding of the nature of these
IR singularities demands an embedding into
the UV-complete string or M-theory, which is beyond our current knowledge.
We hope to return to this problem in future.

   Although in this work we have been focusing on supersymmetric
domain-wall solutions, it is also worthwhile to study other types of
solutions, both supersymmetric and non-supersymmetric. In particular,
there may be new scalar hairy AdS black-hole solutions within the
$SO(3)_D\times SO(3)_R$-invariant sector, which would be the
holographic duals of
conformal field theories at non-zero temperature. The diverse AdS
scalar hairy black holes may bring some new understanding of the phase
structures of the dual field theory. In fact, the
$SO(3)_D\times SO(3)_R$-invariant sector can include
an $\cN=1$ vector multiplet.  This would provide a new arena within which to
look for new $U(1)$-charged AdS dyonic black holes in $SO(8)$
gauged $\cN=8$ supergravity,  in addition to those that were found
within the truncation to the $STU$ supergravity model
\cite{Duff:1999gh,Cvetic:1999xp,Lu:2013ura,Chow:2013gba}, and
the  $SO(3)\times SO(3)$-invariant sector that was studied in
\cite{Cremonini:2014gia,Gauntlett:2009bh}.   The inclusion of both electric
and magnetic charges will lead to a richer structure in the
phase diagrams for  AdS black holes.

    Besides the $\omega$-deformed $SO(8)$ gauging, $\cN=8$ supergravity
allows other possible dyonic gaugings, which also admit supersymmetric
AdS vacua \cite{Gallerati:2014xra}:
 \begin{itemize}
   \item $SO(1,7)$ and $\left[SO(1,1)\times SO(6)\right]\ltimes T^{12}$ supergravities with $\cN=4$
       supersymmetry.
   \item $SO(1,7)$ and $ISO(1,7)$ gauged supergravities with $\cN=3$ supersymmetry.
 \end{itemize}
 Interestingly, the dyonic $ISO(1,7)$ gauged maximal supergravity can arise
from a consistent reduction of massive type IIA
supergravity \cite{Guarino:2015jca}, allowing a stringy interpretation of
the physical consequences of the dyonic gauging, from which many details
of the CFTs dual to the supersymmetric AdS vacua can be deduced. Therefore,
it should be worthwhile to study the supersymmetric domain-wall
solutions, and other type of solutions, in $\cN=8$ supergravities with
different dyonic gaugings.


\section*{Acknowledgements}

  We are grateful to Ofer Aharony, Hadi and Mahdi Godazgar
and Daniel Jafferis for useful discussions. Y.P. was in part
supported by INFN during his stay in the Galileo Galilei Institute
for the workshop ``Holographic
Methods for Strongly Coupled Systems.''  C.N.P. thanks the Mitchell
Foundation for hospitality at the Great Brampton House Workshop on Cosmology
and Strings, April 2015, where part of this work was carried out.
C.N.P. and Y.P. are
supported in part by DOE grant DE-FG02-13ER42020.

\appendix

\section{Branching Rules and Invariant 4-Forms}\label{branching rules}

  The embedding of the $SO(3)_D \times SO(3)_R$ that we are considering in
this paper into $SO(8)$ can be described via the chain of embeddings
\begin{equation}
\begin{array}{lllllllll}
&&SO(8)\rightarrow SO(3)\times SO(5)\rightarrow SO(3)\times SO(4)
\rightarrow SO(3)\times SO(3)_L \times SO(3)_R \rightarrow
SO(3)_D \times SO(3)_R\ ,\nonumber\\
&&35_s\rightarrow (3,10)+(1,5)\rightarrow(3,6)+\ldots(1,1)+\ldots\rightarrow  (3,3,1)+\ldots(1,1,1)\ldots\rightarrow(1,1)+\ldots(1,1)\ldots
\end{array}
\end{equation}
The branching of $35_c$ goes the same way. Starting with
$SO(3)\times SO(5)$, we have
\begin{equation}
\begin{array}{llllll}
dx^1\wedge dx^2 \wedge dx^3 \wedge dx^{\hat{a}}&\in&(1,5)& \\
dx^i\wedge dx^j \wedge dx^{\hat{a}} \wedge dx^{\hat{b}}&\in&(3,10)&\text{with }i=1,2,3\in SO(3)\quad\text{and }\quad\hat{a}\, \hat{b}=4\ldots8\in SO(5).
\end{array}
\end{equation}
Reducing to $SO(3)\times SO(4)$ gives
\begin{equation}
\begin{array}{llllll}
dx^1\wedge dx^2 \wedge dx^3 \wedge dx^{8}&\in&(1,1)& \\
dx^i\wedge dx^j \wedge dx^{a} \wedge dx^{b}&\in&(3,6)&\text{with }
i=1,2,3\in SO(3)\quad\text{and }\quad a\, b=4\ldots7\in SO(4).
\end{array}
\end{equation}
The decomposition under $SO(3)\times SO(3)_L \times SO(3)_R$ gives
\begin{equation}
\begin{array}{llllll}
dx^1\wedge dx^2 \wedge dx^3 \wedge dx^{8}&\in&(1,1,1)& \\
\epsilon^i_{jk}\eta^{\hat{i}}_{ab} dx^i\wedge dx^j \wedge dx^a \wedge
 dx^{b}&\in&(3,3,1)&\text{with }\hat{i}=1,2,3\in SO(3)_L,
\end{array}
\end{equation}
where $\eta^{\hat{i}}_{ab}$ are the 't Hooft matrices.

Finally, the reduction to $SO(3)_D\times SO(3)_R$, where $SO(3)_D$ is
the diagonal in $SO(3)\times SO(3)_L$, gives
\begin{equation}
\begin{array}{llllll}
dx^1\wedge dx^2 \wedge dx^3 \wedge dx^{8}&\in&(1,1)& \\
\epsilon^i_{jk}\eta^{i}_{ab} dx^i\wedge dx^j \wedge dx^a \wedge dx^{b}&\in&(1,1)&\text{with }i=1,2,3\in SO(3)_D
\end{array}
\end{equation}
Note that $i$ is now taken to be an $SO(3)_D$ index, and has been contracted.
The 8-dimensional Hodge dual forms are also invariant, corresponding to
scalars that will be retained.  They can be written as
\begin{equation}
\begin{array}{llllll}
\eta^{i}_{ab}\eta^{i}_{cd} dx^a\wedge dx^b \wedge dx^c \wedge dx^{d}=dx^4\wedge dx^5 \wedge dx^6 \wedge dx^7&& \\
\eta^{i}_{ab}dx^i\wedge dx^a \wedge dx^b \wedge dx^{8}&&&
\end{array}
\end{equation}

We take the 't Hooft matrices to be given by
\begin{equation}
\eta^1=
\left(
\begin{array}{cccc}
 0 & -1 & 0 & 0 \\
 1 & 0 & 0 & 0 \\
 0 & 0 & 0 & 1 \\
 0 & 0 & -1 & 0 \\
\end{array}
\right),\quad
\eta^2=\left(
\begin{array}{cccc}
 0 & 0 & -1 & 0 \\
 0 & 0 & 0 & -1 \\
 1 & 0 & 0 & 0 \\
 0 & 1 & 0 & 0 \\
\end{array}
\right),\quad
\eta^3=\left(
\begin{array}{cccc}
 0 & 0 & 0 & -1 \\
 0 & 0 & 1 & 0 \\
 0 & -1 & 0 & 0 \\
 1 & 0 & 0 & 0 \\
\end{array}
\right)\,.
\end{equation}
These transform as the $\mathbf{3}$ of $so(3)_L$, whose generators
are chosen to be
\bea
B_1=-\frac{1}{2}(R^{01}-R^{23}),\quad
B_2=-\frac{1}{2}(R^{02}-R^{31}),\quad B_3=-\frac{1}{2}(R^{03}-R^{12}).
\eea
The 't Hooft matrices are invariant under $so(3)_R$, whose generators
are chosen to be
\bea
A_1=\frac{1}{2}(R^{01}+R^{23}),\quad A_2=\frac{1}{2}(R^{02}+R^{31}),
\quad A_3=\frac{1}{2}(R^{03}+R^{12}),
\eea
where the $R^{ij}$ are the $so(4)$ generators,
with $(R^{ij})_{ij}=-(R^{ij})_{ji}=1$, and all other elements equal to zero.

\section{Conventions}\label{conventions}

The 8D gamma matrices admits a real representation:
\begin{eqnarray}
\Gamma^1&=&\sigma_2\otimes\sigma_2\otimes\sigma_1\otimes\sigma_0 ,\qquad
\Gamma^2=\sigma_2\otimes\sigma_3\otimes\sigma_0\otimes\sigma_2 ,\nn\\
\Gamma^3&=&\sigma_2\otimes\sigma_0\otimes\sigma_2\otimes\sigma_3 ,\qquad
\Gamma^4=\sigma_2\otimes\sigma_0\otimes\sigma_2\otimes\sigma_1 ,\nn\\
\Gamma^5&=&\sigma_2\otimes\sigma_1\otimes\sigma_0\otimes\sigma_2 ,\qquad
\Gamma^6=\sigma_2\otimes\sigma_2\otimes\sigma_3\otimes\sigma_0 ,\nn\\
\Gamma^7&=&-\sigma_2\otimes\sigma_2\otimes\sigma_2\otimes\sigma_2 ,\quad~
\Gamma^8=\sigma_1\otimes\sigma_0\otimes\sigma_0\otimes\sigma_0 ,\nn\\
\Gamma^9&=&\Gamma^1\dots\Gamma^8=-\sigma_3\otimes\sigma_0\otimes\sigma_0
\otimes\sigma_0,
\label{Gamma}
\end{eqnarray}
where $\sigma_1$, $\sigma_2$ and $\sigma_3$ are the standard Pauli matrices,
and $\sigma_0$ denotes the $2\times 2$ identity matrix.
The gamma matrices are all block off-diagonal:
\be
\Gamma^i=\left(\begin{array}{cc}
0& \hat{\Gamma}^{i}_{I\alpha}\\
(\hat{\Gamma}^i)^{\rm T}_{\alpha I} & 0 \\
\end{array}\right).
\ee
The $8\times8$ matrices $\hat{\Gamma}^i_{I\alpha}$ are the triality tensors,
which map between the three 8-dimensional representations of
$SO(8)$. Note that since $\hat{\Gamma}^8_{I\alpha}=\delta_{I\alpha}$,
the $I$ and $\alpha$  indices are equivalent under $SO(7)$.

The $SO(7)_+$ invariant tensor $C_{+\alpha\beta\rho\lambda}$ is proportional to $ \sum_{i,j=1,7}\hat{\Gamma}^{ij{\rm T}}_{[\alpha\beta}\hat{\Gamma}^{ij{\rm T}}_{\rho\lambda]}$, whilst The $SO(7)_-$ invariant tensor $C_{-IJKL}$ is proportional to $\sum_{i,j=1,7}\hat{\Gamma}^{ij}_{[IJ}\hat{\Gamma}^{ij}_{KL]}$.

   In the construction of the internal metric on the deformed 7-sphere
when calculating the lifting from four to eleven dimensions, it is convenient
to use a parameterisation of the eight Euclidean coordinates $X^I$
describing the embedding of the unit $S^7$, with $X^I X^I=1$, that is
adapted to the symmetries of the system.  We find that it is
convenient to write
\bea
X^1&=& \sin\chi\, \cos\xi\, \cos\tilde\theta\, \cos\varphi_1\,,\quad
X^2= \sin\chi\, \cos\xi\, \cos\tilde\theta\, \sin\varphi_1,\nn\\
X^3&=& \sin\chi\, \cos\xi\, \sin\tilde\theta\, \cos\varphi_2\,,\quad
X^4= \sin\chi\, \cos\xi\, \sin\tilde\theta\, \sin\varphi_2,\nn\\
X^5&=& \sin\chi\, \sin\xi\, \sin\theta\, \cos\varphi_3\,,\quad
X^6= \sin\chi\, \sin\xi\, \sin\theta\, \sin\varphi_3,\nn\\
X^7&=& \sin\chi\, \sin\xi\, \cos\theta\,,\quad\quad\quad~~ X^8 =\cos\chi.
\label{newxpara}
\eea
The metric $d\Omega_7^2 =\delta_{AB}dX^A dX^B$ on the unit $S^7$
is now given by
\be
d\Omega_7^2 = d\chi^2 + \sin^2\chi\, d\Omega_6^2,
\ee
with the metric $d\Omega_6^2$ on the unit $S^6$ given in terms of metrics
$d\Omega_2^2$ and $d\Omega_3^2$ on a unit $S^2$ and a unit $S^3$ by
\bea
d\Omega_6^2 &=& d\xi^2 + \sin^2\xi\, d\Omega_2^2 +
                \cos^2\xi\, d\Omega_3^2,\nn\\
d\Omega_2^2 &=& d\theta^2 + \sin^2\theta\, d\varphi_3^2\,,\qquad
d\Omega_3^2 = d\tilde\theta^2 + \cos^2\tilde\theta\, d\varphi_1^2 +
     \sin^2\tilde\theta\, d\varphi_2^2.
\eea

    As a check of our uplift procedure, we can present explicitly
the internal metric for the simpler $G_2$-invariant sector, which
can be obtained from the $SO(3)_D\times SO(3)_R$ invariant sector by
setting $\phi_1=\phi_2$ and $\sigma_1=-\sigma_2$. Using the uplift
ansatz (\ref{metans}), we derive the internal metric
for the $G_2$ invariant sector
\be
ds_7^2=\Delta^{-1}\Big(a^{-3}d\chi^2+
\fft{\sin^2\chi}{a\, (a^2 \cos^2\chi+b^2 \sin\chi^2)}\, d\Omega_6^2
\Big),
\label{G2uplift}
\ee
where
\bea
a&=& \cosh \phi _1-\cos \sigma _1 \sinh \phi _1,\nn\\
b&=& \cosh \phi _1+\cos \sigma _1 \sinh\phi _1,\nn\\
\Delta&=& a^{-1} \left(a^2 \cos^2\chi+b^2 \sin^2\chi\right)^{-2/3}.
\eea
This result matches with the one given in \cite{Ahn:2001kw}.

\end{document}